# Model-Free Fast Frequency Support of Wind Farms for Tracking Optimal Frequency Trajectory

Yubo Zhang, *Student Member, IEEE,* Songhao Yang, *Member, IEEE,* Zhiguo Hao, *Senior Member, IEEE,* Baohui Zhang, *Fellow, IEEE*

***Abstract*—The fast frequency support (FFS) towards frequency trajectory optimization provides a system view for the frequency regulation of wind farms (WFs). However, the existing frequency trajectory optimization-based FFS generally relies on the accurate governor dynamics model of synchronous generators (SGs), which aggrandizes the difficulty of controller implementation. In this paper, a proportional-integral (PI) based FFS of WFs is designed for tracking the optimal frequency trajectory, which gets rid of the dependence on the governor model. Firstly, the prototypical PI-based FFS of WFs is proposed and its feasibility for tracking the optimal frequency trajectory is analyzed and demonstrated. Then, based on the "frequency-RoCoF" form of the optimal frequency trajectory, a more practical PI controller is constructed, avoiding the time dependence of the prototypical PI controller. Besides, an adaptive gain associated with PI parameters is designed for multi-WF coordination. Finally, the validity of the proposed method is verified in both the single-WF system and the multi-WF system.**

## I. Introduction

With the large-scale converter interfaced generators (CIGs) connected to the grid, the inertia of the power system is seriously diminished [1]. Low inertia significantly increases the over-limit risk of the frequency and threatens the safety of power consumption. Hence, excavating the potential resources to improve frequency safety has attracted extensive attention.

To meet this urgent demand for power systems, extensive research has been carried out on the frequency support of wind farms (WFs). For the existing frequency support control of WFs, the energy sources mainly include de-loading [2], energy storage [3], and rotor kinetic energy (KE). Both de-loading and energy storage enable the WF to provide long-term frequency support, as a certain portion of power is reserved. However, the energy storage devices require heavy investment whereas the de-loading will cause economic losses of wind generation. So, the WF owners generally prefer frequency support control by releasing the KE, because the WF can maintain the maximum power point tracking (MPPT) operation mode under normal conditions. Given the limited KE, the WF operating in MPPT mode can only provide short-term frequency support.

In terms of the KE-based frequency support control of WFs, the currently mainstream strategies can be roughly divided into two types, namely the virtual inertia control (VIC) and the stepwise inertia control (SIC). The core idea of VIC is to generate an additional power proportional to the rate of change of frequency (RoCoF) and the frequency excursion [4]. Hence, control gain tuning is the key issue of VIC, as it reveals a trade-off between VIC performance and the safe operation of wind turbines (WTs). Based on the prototypical VIC in [4], the VIC with adaptive gains has been studied in [5-9]. The releasable KE-based method was proposed in [5], which configured larger control gains for the WT with more releasable KE. To prevent the possible over-deceleration, the adaptive gain related to real-time rotor speed was proposed in [6-9]. By introducing rotor speed into the control gain expressions, the frequency support is attenuated with the deceleration of WT to ensure its safety. Considering the correlation of the frequency support demand with the severity of power deficit, the frequency information is further introduced into the design of control gains, disturbance information-based adaptive gain strategy was proposed in [10-13], such as frequency [10], frequency deviation [11], RoCoF [12], and both frequency and RoCoF [13], aiming at improving the adaptability of VIC to different disturbance sizes.

The SIC provides frequency support by directly regulating the output power of WFs as the predefined shape. In [14], the SIC was first proposed to provide a fixed support power and then maintain it for a period of time, which effectively alleviates the initial unbalanced power and improves the frequency nadir. However, the secondary frequency drop (SFD) is inevitable as the WTs will produce a large power plunge for speed recovery. In view of this, some improved SIC strategies are proposed by developing a superior power reference curve. In [15-17], the magnitude and duration time of the overproduced power were optimized. Besides, different shapes of power reference curves were designed in [18-22], such as straight curves in the time-power plane [18], straight curves [19-21], and exponential curves [22] in the speed-power plane. By optimizing the output power reference, the SFD can be effectively suppressed.

However, both VIC and SIC lack a systematic perspective, that is, how to provide better frequency support for the power system based on limited KE. Considering the essential intention of the fast frequency support (FFS) of WFs is to improve the system frequency safety, recent research attempts to explore the system-oriented FFS of WFs from the perspective of optimizing frequency trajectories. To maximize the frequency nadir, the model predictive control (MPC) was adopted in [23] to regulate

This work was supported by National Natural Science Foundation of China (52007143), China Postdoctoral Science Foundation (2021M692526) and Key Research and Development Program of Shaanxi(2022GXLH-01-06)
Y. Zhang, S. Yang, Z. Hao and B. Zhang are with State Key Laboratory of Electrical Insulation and Power Equipment, Xi'an Jiaotong University, Xi'an, China (e-mail: zyb970305@stu.xjtu.edu.cn, {songhaoyang, zhghao, bhzhang}@xjtu.edu.cn).

the active power of WFs. However, given the limited prediction interval of MPC, the frequency trajectory in the whole primary frequency regulation (PFR) process cannot be sensed at one time. In [24] and [25], a first-order frequency trajectory satisfying both the RoCoF and frequency nadir constraints was proposed empirically and intuitively. Then, its optimality was numerically verified by constructing and solving the optimal control model in [26]. However, the numerical solution is less rigorous, and the particle swarm optimization-based solver is too complicated to balance the accuracy and computational efficiency. From the view of frequency regulation energy, the universality of the optimal frequency trajectory was strictly proved in [27], which was independent of the system model and power disturbance. Based on that, a series of FFS strategies have been designed to achieve the optimal frequency trajectory. In [24], the time-domain optimal frequency trajectory was applied in constructing the damping-inertia control for the FFS of the inverter. Taking the optimal frequency trajectory as the output of the frequency response model, the FFS was reversely derived with energy storage [25] and WFs [26-27] as controlled plants. However, these model-based FFS schemes rely on complex nonlinear governor models and detailed parameters. While theoretically sound, these methods may be overly idealistic when applied to large-scale power systems.

Taking the optimal frequency trajectory as the control goal, this paper proposes a PI-based FFS of WFs dispensing with the governor modeling. The main contributions are:

1) The prototype of PI-based FFS is first constructed, and its viability for tracking the optimal frequency trajectory is analyzed theoretically. Besides, the analytical expression of PI parameters is given. The model-free property makes the proposed method dispense with governor modeling, which significantly simplifies its implementation.
2) The time-independent PI-based FFS is proposed based on the "frequency-RoCoF" form of the optimal frequency trajectory. By reconstructing the generation method of the reference signal, the time dependence of the prototypical PI controller is eliminated, and the robustness is improved.
3) An adaptive gain strategy is further designed for multi-WF coordination. This strategy adaptively adjusts the gain of the PI controller according to the kinetic energy of the WF, so as to reasonably allocate FFS tasks among WFs. Since only WF local information is needed, it achieves the multi-WF coordination without extra communication demands.

The rest of the paper is organized as follows. Section II briefly introduces the optimal frequency trajectory and the model-based implementation. Section III shows the prototype of the PI controller for achieving the optimal frequency trajectory and the PI parameters design. Section IV derives the "frequency-RoCoF" form of optimal frequency trajectory and proposes the time-independent PI-based FFS of WFs. The proposed method is verified in Section V. Finally, Section VI draws a conclusion.

## II. System-view Optimal Frequency Trajectory

### A. System Frequency Dynamics

By aggregating the rotor of all synchronous generators (SGs), one obtains the following average frequency dynamics of the power system.

$$2H\Delta \dot{f}(t) = \Delta P_m(t) - \Delta P_e(t) - D_f \Delta f(t) \tag{1}$$

where $H$ is the system inertia; $D_f$ is the frequency damping coefficient that is related to the frequency-dependent nature of loads; $f$ is the system frequency; $\Delta P_m$ and $\Delta P_e$ denote the mechanical power and electrical power of the power system.

During the process of PFR, the mechanical power $\Delta P_m$ can be approximated as

$$\Delta P_m = \Delta f(s) * G_{gov}(s) \tag{2}$$

where $G_{gov}(s)$ is the model of the governor and turbine system.

Taking the FFS of WFs into consideration, the electrical power $\Delta P_e$ under a power deficit can be expressed as

$$\Delta P_e(t) = P_d - \Delta P_{WF}(t) \tag{3}$$

where $P_d$ is the power deficit, which is supposed to be a fixed value in this paper; $\Delta P_{WF}$ is the frequency response of WFs.

In this paper, WFs are supposed to operate in the MPPT mode normally and only provide FFS by releasing the kinetic energy. Hence, the steady-state frequency excursion of the PFR $\Delta f_{ss}$ is unaffected, which is given in (4) according to [28].

$$\Delta f_{ss} = -P_d / K_g \tag{4}$$

where $K_g = D_f + 1/R$ is the compositive gain of PFR; $R$ is the system droop factor.

### B. Wind Farm Modeling

As the capacity of the actual WT is very limited, the term "wind farm" is used as the aggregation of WTs to reflect the scale effect in this paper. For the convenience of analysis, we assume that all WTs in the WF are the same, so the WF can be regarded as an equivalent WT with a large capacity.

The aerodynamic power captured by the turbine blades is

$$P_a = 0.5\rho\pi R^2 v_w^3 C_p(\lambda, \beta) \tag{5}$$

where $\rho$ is the air density; $R$ is the turbine radius; $v_w$ is the wind speed; $\lambda$ is the tip-speed ratio; $\omega_r$ is the rotor speed of the WT; $\beta$ is the pitch angle; $C_p$ is the aerodynamic factor.

The mechanical power is transmitted to the generator of the WT by a shaft, whose dynamics can be depicted as follows

$$\frac{d\omega_r}{dt} = \frac{1}{J_{WT}}\left(\frac{P_a}{\omega_r} - \frac{P_{WT}}{\omega_r}\right) \tag{6}$$

where $J_{WT}$ is the inertia of the WT.

In this paper, the active power of the WT under the proposed FFS scheme is set as

$$P_{WT} = P_{WT0} + \Delta P_{WT} \tag{7}$$

where $P_{WT0}$ is the steady-state power of WT before disturbance; $\Delta P_{WT}$ is the additional power of WT for frequency support.

Then, the frequency support power of the wind power system can be further calculated as

$$\Delta P_{WF} = \sum_{i=1}^{N_{WF}} \sum_{j=1}^{N_{WT,i}} \Delta P_{WT,ij} \tag{8}$$

where $N_{WF}$ is the number of WFs; $N_{WT,i}$ is the number of WTs in the $i$-th WF; $\Delta P_{WT,ij}$ is the support power of the $j$-th WT in the $i$-th WF.

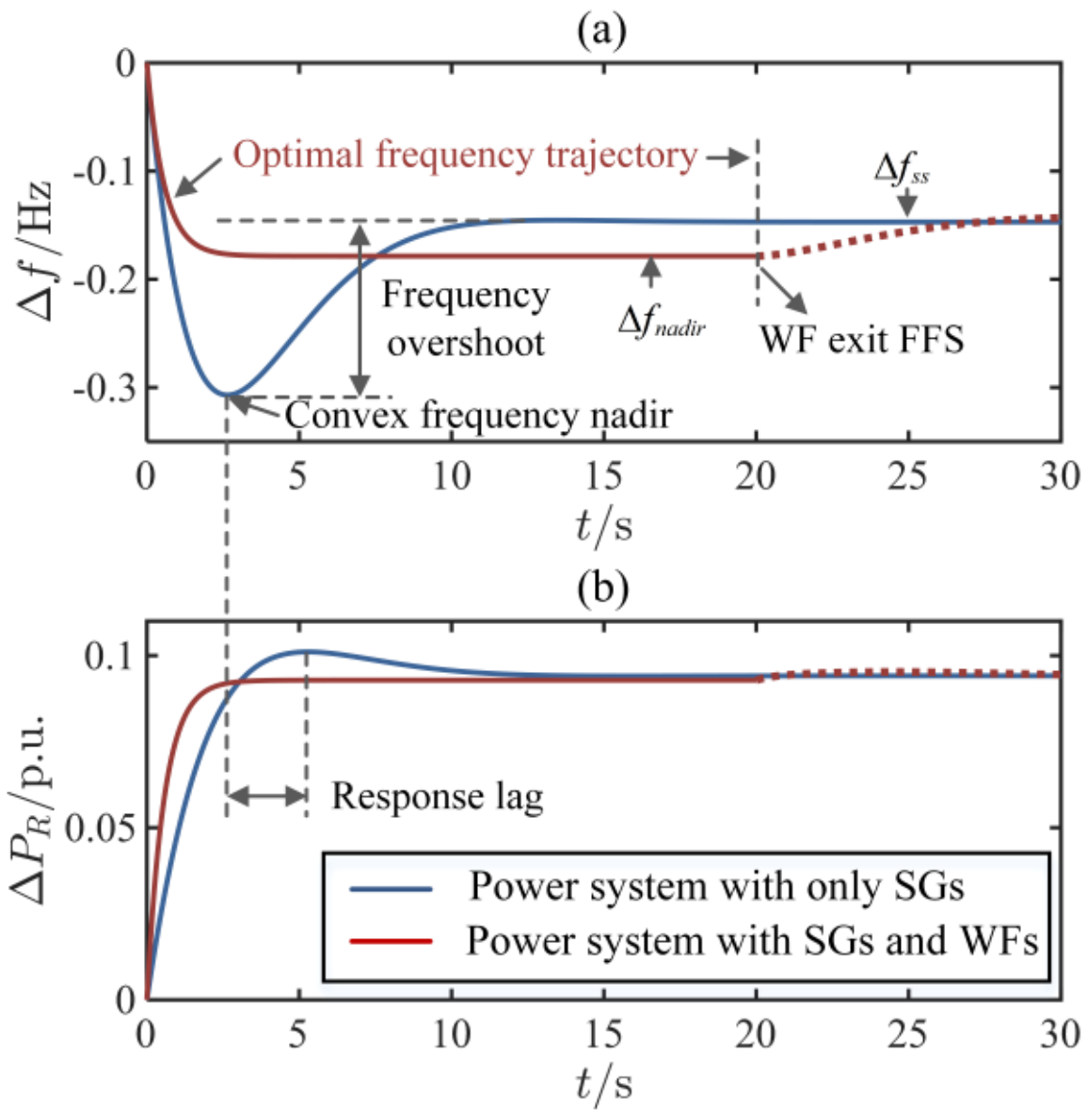

Fig. 1 Comparison of the conventional and optimal frequency trajectory. (a) Frequency trajectory. (b) Frequency regulation power.

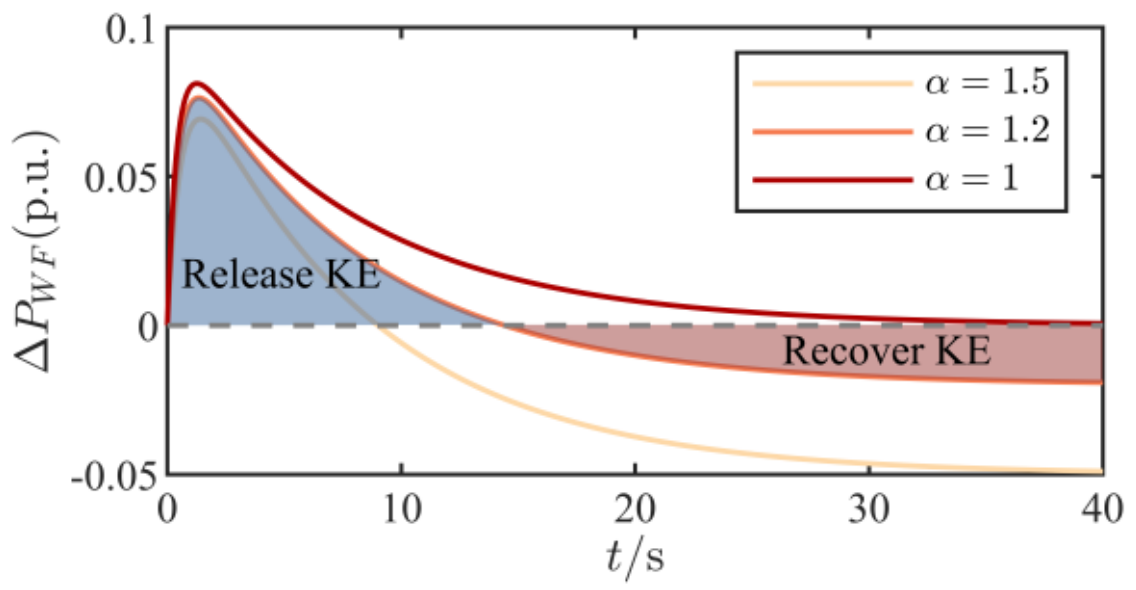

Fig. 2 Support power of WFs under different values of $\alpha$ .

## *C. Optimal Frequency Trajectory*

Thinking outside of imitating the frequency response of SGs, the frequency trajectory optimization-oriented FFS of WFs has attracted much attention recently. The frequency trajectory shown in (9) is widely used in [24-27], which can be regarded as an attractive form of the optimal frequency trajectory. That is, **the frequency drops exponentially to the frequency nadir and then maintains until the FFS scheme exits**, as shown by the solid part of the red curve in Fig. 1(a).

$$\Delta f_{opt}(t) = A_f(1 - e^{-t/T_f}) \tag{9}$$

The optimal frequency trajectory satisfies the following two conditions, namely the initial RoCoF and the frequency nadir.

$$\begin{cases} \Delta \dot{f}(0) = A_f / T_f = -P_d / 2H \\ \Delta f(\infty) = A_f = \Delta f_{nadir} \end{cases} \tag{10}$$

where $\Delta f_{nadir}$ is the excursion of the frequency nadir.

According to (10), the parameters in (9) can be derived as

$$A_f = -\frac{\alpha P_d}{K_g}, \quad T_f = \frac{2\alpha H}{K_g} \tag{11}$$

where $\alpha = \Delta f_{nadir} / \Delta f_{ss}$ denotes the ratio of the frequency nadir to the steady-state frequency, which can be specified according to operation demands. Fig. 2 shows the support power of WFs under different values of $\alpha$ . It can be seen that the smaller of $\alpha$ , the slower the KE recovery ( $\Delta P_{WF} < 0$ ), and the over-deceleration risk of WFs will increase. If $\alpha = 1$, the recovery time of KE tends to infinity. The overlong recovery process will threaten the safe operation of WFs. Hence, it is recommended to be no less than 1.1. The selection of $\alpha$ is not the focus of this paper, so there is no further discussion.

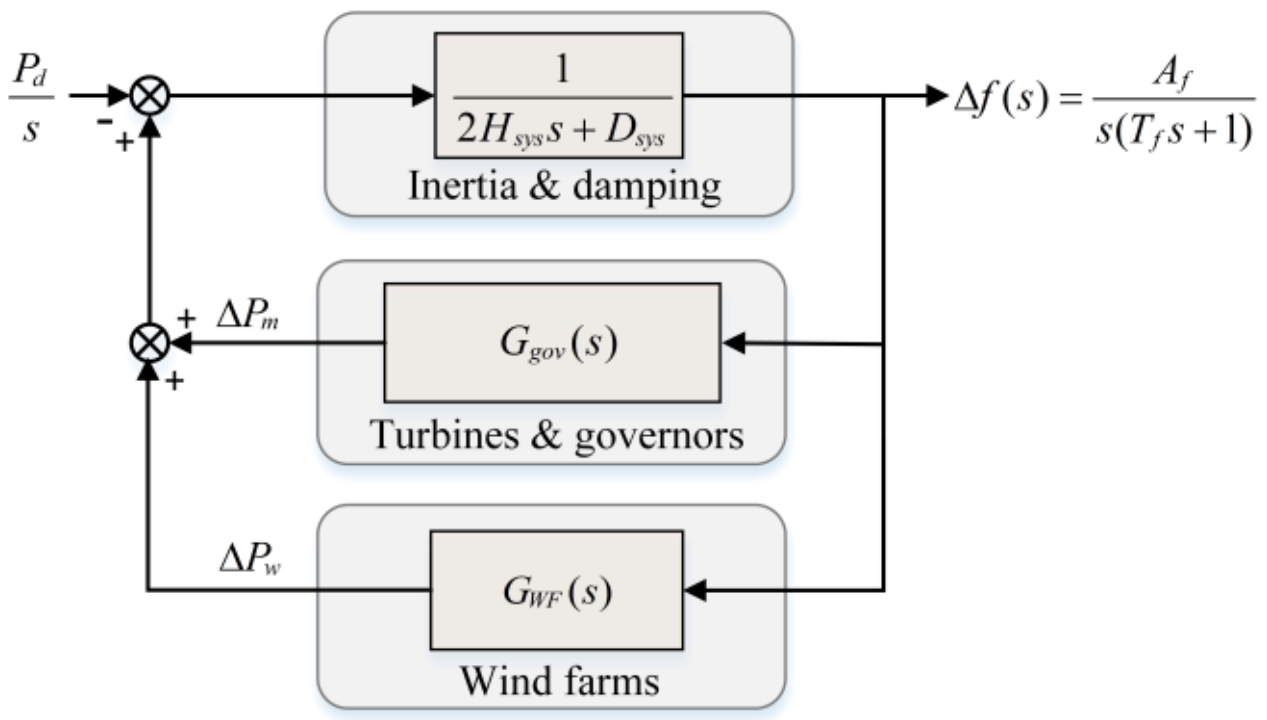

Fig. 3 Schematic diagram of the model-based FFS of WFs in [26] and [27].

Combined with Fig. 1, the optimality of the above frequency trajectory is further elaborated. Firstly, the optimal trajectory eliminates the convex frequency nadir, which is essentially the frequency overshoot caused by the response lag of the governor system of SGs. Hence, the optimal frequency trajectory means an effective coordination between SGs and WFs, that is, the FFS of WFs compensates for the response lag of SGs. Secondly, the optimal frequency trajectory can maximize the frequency nadir with the same frequency regulation energy. This merit of the optimal frequency trajectory has been strictly proved in our previous work [27]. Finally, the shape of the optimal frequency trajectory is very simple, and its key transient indices can be calculated algebraically, as shown in (10). This property can significantly simplify the frequency security assessment as there is no need for the dynamic simulations to get the convex frequency nadir.

## *D. Model-based FFS Scheme of WFs And Its Shortcomings*

Taking the optimal frequency trajectory as the desired output under the power deficit, the model-based FFS control schemes is reversely derived from the system frequency model in [26] and [27], with the principle shown in Fig. 3. Specifically, the *s*-domain optimal frequency trajectory can be obtained by *Laplace* transform to (9), as follows.

$$\Delta f_{opt}(s) = \frac{A_f}{s(T_f s + 1)} \tag{12}$$

Then, the FFS of WFs can be reversely constructed as

$$G_{WF}(s) = -G_{gov}(s) - K_w \tag{13}$$

where $K_w = D_f - K_g / \alpha$ is a constant gain; $G_{gov}(s)$ involves the governor model of all SGs.

As shown in (13), the model-based FFS of WFs relies on the accurate governor dynamics of all SGs, which usually includes complicated dynamical equations and nonlinear characteristics such as saturation and clipping. For a large-scale power system with numerous SGs, the workload of modeling all the governor dynamics is heavy, and model errors will inevitably result in control performance degradation.

## III. Prototype of PI Controller for FFS of WFs

To get rid of the governor modeling, a PI-based FFS of WFs is proposed in this paper. The core idea is to construct a PI controller to regulate the output power of WFs, with optimal frequency trajectory tracking as the control goal. The model-free property of the PI controller endows the proposed method free from governor modeling, so as to improve its application potential in actual power systems. The fundamental principles are introduced as follows.

### *A. Control Prototype*

Given the flexible control of WFs, the system frequency can theoretically be shaped into the desired trajectory by regulating the output power of WFs. Therefore, if the desired frequency trajectory is known, the following PI-based FFS of WFs can be constructed to force the system frequency to track the reference frequency, as follows.

$$\Delta P_{WF}(t) = K_P * e_f(t) + K_I * \int e_f(t)dt \tag{14}$$

where $e_f(t) = \Delta f_{ref}(t) - \Delta f_{act}(t)$, which denotes the tracking error; $\Delta f_{ref}$ is the reference frequency; $\Delta f_{act}$ is the actual frequency; $K_P$ and $K_I$ are coefficients of the PI controller.

Substituting (14) into (1), one obtains

$$\Delta f_{act}(s) = \frac{-\frac{P_d}{s} + G_{PI}(s)\Delta f_{ref}(s)}{G_{sys}(s) + G_{PI}(s) - G_{gov}(s)} \tag{15}$$

where $G_{sys}(s) = 2Hs + D_f$, $G_{PI}(s) = K_P + K_I/s$.

In this paper, the desired system frequency is exactly the optimal frequency trajectory. That is

$$\Delta f_{ref}(s) = \Delta f_{opt}(s) \tag{16}$$

Based on (12), the power deficit and the optimal frequency trajectory should satisfy the following equation.

$$-\frac{P_d}{s} = (2Hs + K_g^*)\Delta f_{opt}(s) \tag{17}$$

where $K_g^* = K_g/\alpha$.

Combining (15), (16), and (17), the frequency response can be further expressed as

$$\Delta f_{act}(s) = \left[1 + G_R(s)\right]\Delta f_{opt}(s) \tag{18}$$

where

$$G_R(s) = \frac{K_g^* - D_f + G_{gov}(s)}{G_{sys}(s) + G_{PI}(s) - G_{gov}(s)} \tag{19}$$

According to (18), the premise for the system frequency to track the optimal frequency trajectory is that the influence of $G_R(s)$ can be ignored. Mathematically, it is equivalent to

$$\left|G_R(s = j\omega)\right| \ll 1, \omega_{low} \le \omega \le \omega_{up} \tag{20}$$

where $\omega_{low}$ and $\omega_{up}$ are the lower bound and upper bound of the spectrum of $\Delta f_{opt}$. In general, let $\omega_{low} = 0$.

### *B. Spectrum of the Optimal Frequency Trajectory*

According to (20), it is necessary to study the spectrum of $\Delta f_{opt}$. The fast Fourier transform (FFT) can be performed by Matlab to get the spectrum of optimal frequency trajectory. The amplitudes and the frequencies of all FFT components are denoted as $\{A_0, A_1, \cdots, A_{(N_S/2)}\}$ and $\{\omega_0, \omega_1, \cdots, \omega_{(N_S/2)}\}$, where $N_S$ is the sampling length. To determine the upper bound $\omega_{up}$, the following criterion is adopted. That is, once the ratio of cumulative spectral energy to total spectral energy exceeds the preset threshold, the maximum frequency is recorded as the upper bound. That is

$$\sum_{i=0}^{k} A_i^2 \Big/ \sum_{i=0}^{N_S/2} A_i^2 \ge 1 - \varepsilon \tag{21}$$

$$\omega_{up} = \omega_k \tag{22}$$

where $\varepsilon$ is a tolerance coefficient, which is taken as $10^{-4}$ in this paper, indicating that the cumulative spectral energy within $[\omega_{low}, \omega_{up}]$ accounts for 99.99% of the total spectral energy.

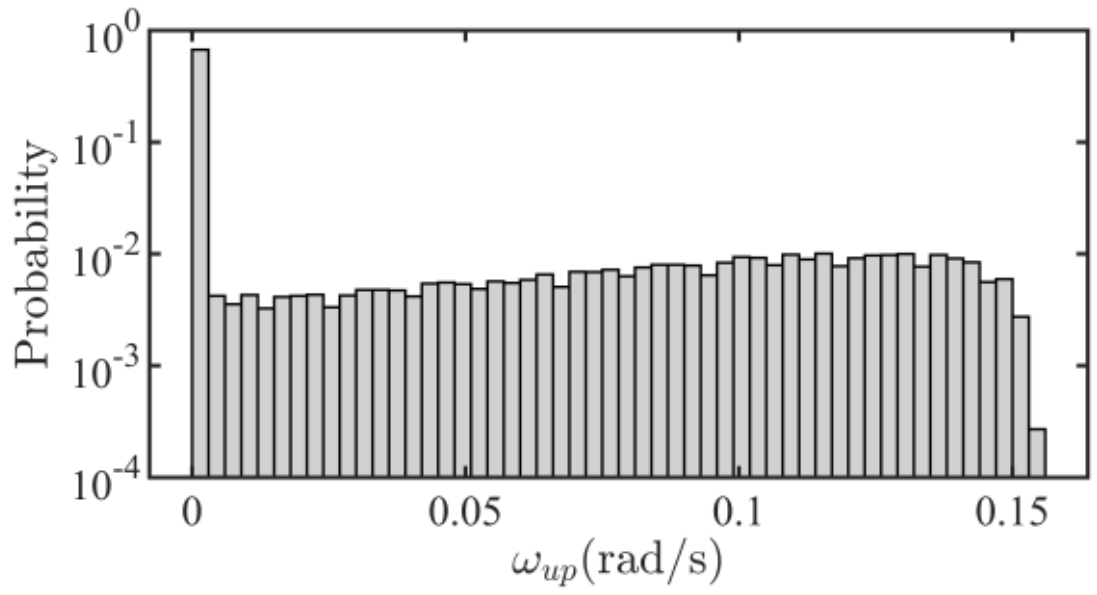

Fig. 4 Statistical result of spectral upper bound $\omega_{up}$.

TABLE I
Parameter Range for Sampling

| Parameters | Lower bound | Upper bound |
|---|---|---|
| Power deficit $P_d$ (p.u.) | 0.01 | 0.5 |
| Ratio coefficient $\alpha$ | 1 | 5 |
| System inertia time $H$ (s) | 0.1 | 20 |
| System damping factor $D_f$ | 0 | 15 |
| Governor time constant $T_g$ (s) | 0 | 20 |
| Aggregated droop factor $R$ | 0.01 | 1 |

Eq. (9) indicates that the shape of the optimal frequency trajectory is impacted by the system parameters, thus affecting the spectral upper bound. Given this, the spectral upper bound under different system parameters is further verified based on the criterion in (21). The system parameter ranges for sampling are listed in TABLE I, which are larger than those of the actual power system [29]. Then, the uniform sampling within the range from TABLE I is performed and the spectral upper bound $\omega_{up}$ is then calculated. A total of 100,000 samples were taken and the statistical results are shown in Fig. 4. The sampling results show that the spectral upper bound generally does not exceed 0.15rad/s under various values of system parameters. Therefore, denoting $\omega_{up}^{max} = 0.15\,\text{rad/s}$, $\omega_{up}^{max}$ is large enough to depict the spectral upper bound of the optimal frequency trajectory under different power systems.

### *C. Design of PI Parameters*

The design principle of PI parameters is to make (20) hold. To facilitate analysis, the following simplified governor model is adopted in the derivation process.

$$G_{gov}(s) = -\frac{1}{R(1 + T_g s)} \tag{23}$$

where $T_g$ is the time constant of the governor system.

Substituting $s = j\omega$ into (19), $G_R(s)$ can be expressed as

$$G_R(j\omega) = \frac{G_R^{num}(j\omega)}{G_R^{den}(j\omega)} = \frac{a_{num} + jb_{num}}{a_{den} + jb_{den}} \tag{24}$$

where

$$\begin{cases} a_{num} = K_g^* - D_f - \dfrac{1}{R\left[1+\left(\omega T_g\right)^2\right]} \\ b_{num} = \dfrac{\omega T_g}{R\left[1+\left(\omega T_g\right)^2\right]} \\ a_{den} = K_P + D_f + \dfrac{1}{R\left[1+\left(\omega T_g\right)^2\right]} \\ b_{den} = 2H\omega - \dfrac{K_I}{\omega} - \dfrac{\omega T_g}{R\left[1+\left(\omega T_g\right)^2\right]} \end{cases} \tag{25}$$

Combining (20) and (24), one obtains

$$\left|G_R^{num}(j\omega)\right| \ll \left|G_R^{den}(j\omega)\right|, \omega_{low} \le \omega \le \omega_{up} \tag{26}$$

Generally, a number that is an order of magnitude smaller than another number can be considered to be "far less than". So, the qualitative relationship in (26) can be converted into the following quantitative relationship.

$$10\left|G_R^{num}(j\omega)\right| \le \left|G_R^{den}(j\omega)\right|, \omega_{low} \le \omega \le \omega_{up} \tag{27}$$

A sufficient condition for (27) to hold is that both the real part and the imaginary part of $G_R^{num}$ are much smaller than those of $G_R^{den}$, that is

$$\begin{cases} 10\left|a_{num}(j\omega)\right| \le \left|a_{den}(j\omega)\right| \\ 10\left|b_{num}(j\omega)\right| \le \left|b_{den}(j\omega)\right| \end{cases}, \omega_{low} \le \omega \le \omega_{up} \tag{28}$$

Then, the value range of $K_P$ and $K_I$ can be derived as (29) with the detailed derivation process exhibited in Appendix A.

$$\begin{cases} K_P \ge \max\begin{pmatrix} K_g^* - D_f - \dfrac{1}{R\left[1+\left(\omega_{up}^{\max} T_g^{\max}\right)^2\right]} \\ 10\left(K_g - K_g^*\right) \end{pmatrix} \\ K_I \ge 9\dfrac{\omega_{up}^{\max}}{2R} \end{cases} \tag{29}$$

where max() is the function to get the maximum value; $T_g^{\max}$ is the maximum value of $T_g$. According to TABLE I, we take $T_g^{\max} = 20\text{s}$ in this paper. It should be noted that $\omega_{up}^{\max}$ and $T_g^{\max}$ are both constant and do not vary with power systems. This property greatly simplifies the PI parameters design.

Based on the general design principles of PI controller, the value of $K_P$ and $K_I$ should not be too large. So, the minimum $K_P$ and $K_I$ that satisfy (29) are recommended to use, that is

$$\begin{cases} K_{P0} = \min(K_P) \\ K_{I0} = \min(K_I) \end{cases} \tag{30}$$

where min() is the function to get the minimum value; $K_{P0}$ and $K_{I0}$ are the recommended PI parameters.

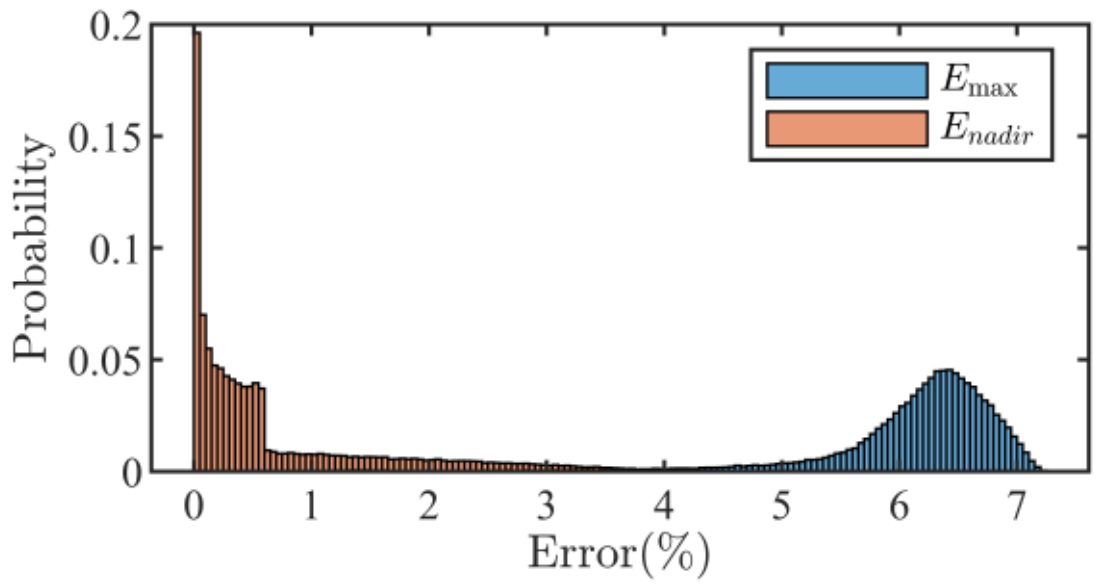

Fig. 5 Error statistics of frequency trajectories.

### D. Performance Verification of the PI Controller

According to the design principle of PI parameters shown in (29), the performance of the PI controller is further verified under various power deficits and power systems. Specifically, the relative parameters are sampled within the range shown in TABLE I, and the corresponding PI parameters are derived by (29). Then, the optimal frequency trajectory shown in (9) and the actual system frequency shown in (18) can be calculated. Finally, the performance of the PI controller is evaluated by the following two relative error indices.

$$\begin{cases} E_{\max} = \max\left(\left|\dfrac{\Delta f_{act} - \Delta f_{opt}}{\Delta f_{opt}}\right|\right) \times \% \\ E_{nadir} = \left|\dfrac{\Delta f_{act,nadir} - \Delta f_{opt,nadir}}{\Delta f_{opt,nadir}}\right| \times \% \end{cases} \tag{31}$$

where $E_{\max}$ denotes the maximum relative error of the actual system frequency trajectory to the optimal frequency trajectory; $E_{nadir}$ denotes the relative error of the actual frequency nadir to that of the optimal frequency trajectory. $\Delta f_{act,nadir}$ and $\Delta f_{opt,nadir}$ are the frequency nadir of the actual frequency and the optimal frequency trajectory, respectively.

Similarly with Section III.B, a total of 100,000 sampling tests were performed, and the statistical results of the above two error indices are shown in Fig. 5. Within a large range of system parameter values, the maximum relative error $E_{\max}$ is less than 8% and the frequency nadir error $E_{nadir}$ is less than 4%, which indicates good consistency between the actual frequency and the optimal frequency trajectory. Hence, the validity of the PI control for tracking the optimal frequency trajectory is verified.

## IV. Proposed PI-based FFS of WFs

In Section III, the theoretical feasibility of PI-based FFS for tracking the optimal frequency trajectory is proved. However, the optimal frequency trajectory in (9) is not appropriate for directly constructing the PI controller, as the time dependence of the reference signal may impair the control robustness. To cope with this problem, the equivalent time-independent form of the optimal frequency trajectory is derived for constructing a practical PI controller for the FFS of WFs.

### A. Time-independent PI Controller for FFS

#### 1) Time-independent optimal frequency trajectory

According to (9), the RoCoF corresponding to the optimal frequency trajectory can be expressed as

$$\Delta\dot{f}_{opt}(t) = -\frac{P_d}{2H} * e^{-t/T_f} \tag{32}$$

Combining (9) and (32), the variable $t$ can be eliminated. Then we obtain the following equivalent form of the optimal frequency trajectory.

$$\Delta\dot{f}_{opt}(t) = -\frac{1}{T_f}\Delta f_{opt}(t) - \frac{P_d}{2H} \tag{33}$$

This equivalent form reveals the dynamic features of optimal frequency trajectory as well as gets rid of the time dependence. That is to say, if the RoCoF and the frequency deviation satisfy (33), the actual frequency is certainly optimal.

*2) Design of time-independent PI controller*

Based on (33), the time-independent PI controller for the FFS of WFs is further designed, as shown in Fig. 6. Specifically, the reference RoCoF is firstly derived from the time-independent optimal frequency trajectory shown in (33) based on the actual frequency, and then the reference frequency is generated by integrating the reference RoCoF. Finally, the deviation between the actual frequency and the reference frequency is fed into the PI controller to generate the additional power command of WFs.

Under the control structure shown in Fig. 6, the additional power of WFs can be derived as

$$\Delta P_{WF} = -\left(1+\frac{1}{T_f s}\right)G_{PI}(s)\Delta f_{act}(s) - \frac{P_d}{2Hs^2}G_{PI}(s) \tag{34}$$

Substituting (34) and (17) into (1), the system frequency dynamics can be expressed as

$$\Delta f_{act}(s) = \left[1+G_R^*(s)\right]\Delta f_{opt}(s) \tag{35}$$

where

$$G_R^*(s) = \frac{K_g^* - D_f + G_{gov}(s)}{G_{sys}(s) + G_{PI}^*(s) - G_{gov}(s)} \tag{36}$$

$$G_{PI}^*(s) = \left(1+\frac{1}{T_f s}\right)G_{PI}(s) \tag{37}$$

It can be seen that $G_R(s)$ and $G_R^*(s)$ have almost the same expression. In Section III C and D, it has been verified that the influence of $G_R(s)$ under the PI parameters selection shown in (29) can be ignored within the spectral range of $\omega \in [0, \omega_{up}^{max}]$. Naturally, if the amplitude of $G_R^*(s)$ is comparable to that of $G_R(s)$ in the spectral range of $\omega \in [0, \omega_{up}^{max}]$, $G_R^*(s)$ can be ignored in (35) likewise.

To verify the validity of the time-independent PI controller, the amplitudes of $G_R(s)$ and $G_R^*(s)$ are compared in the spectral range of $\omega \in [0, \omega_{up}^{max}]$ by sampling, with results shown in Fig. 7. Specifically, equal interval sampling is performed in the spectral range of $\omega \in [0, \omega_{up}^{max}]$. For each $\omega$, a total of 10,000 samplings were performed in the parameter value ranges listed in TABLE I, and then the amplitudes of $G_R(s)$ and $G_R^*(s)$ are calculated. The statistical results in Fig. 7 indicate that $|G_R^*(s)|$ is smaller than $|G_R(s)|$ in the spectral range of $\omega \in [0, \omega_{up}^{max}]$, indicating that $G_R^*(s)$ exerts even less influence on actual frequency than $G_R(s)$. Hence, the item of $G_R^*(s)$ can be ignored in (35), and the proposed control will make the actual system frequency track the optimal frequency trajectory.

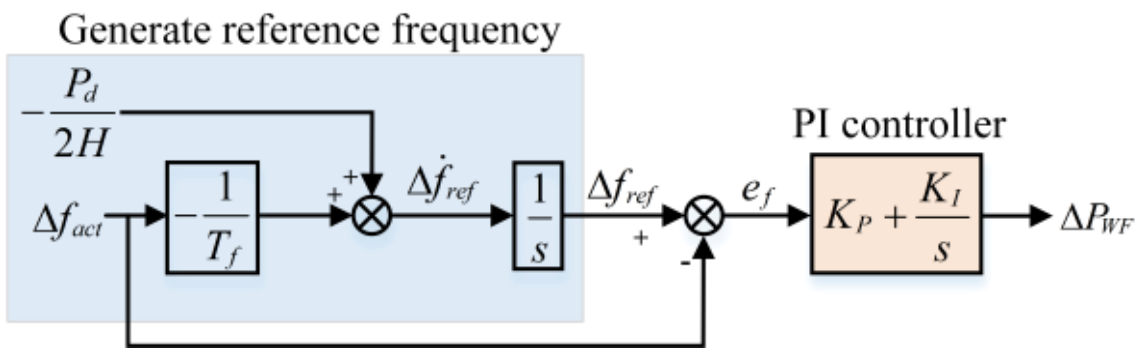

Fig. 6 Time-independent PI controller for the FFS of WFs.

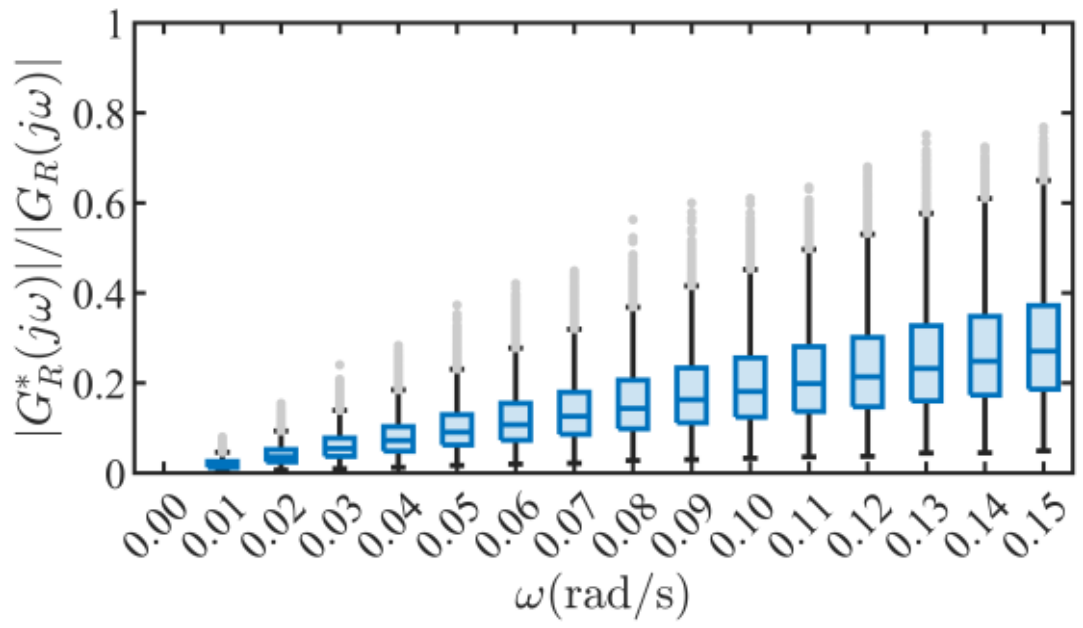

Fig. 7 Statistical results of the amplitude comparison of $G_R^*(s)$ and $G_R(s)$ in the spectral range of $\omega \in [0, \omega_{up}^{max}]$.

*B. Adaptive PI Gains for Multi-WF Coordination*

For the multi-WF system, the FFS task required for tracking the optimal frequency trajectory should be reasonably allocated among all WFs. Given the difference in FFS capability of WFs, the following adaptive gains configuration strategy is designed for multi-WF coordination without extra communication.

Generally, the output amplitude of the PI controller can be regulated by adjusting the PI parameters. Specifically, under the same input, the larger the PI parameters, the larger the output of the PI controller. Hence, the proportion of FFS tasks undertaken by the WF can be controlled by configuring its PI parameters, which is designed as (38) in this paper.

$$\begin{cases} K_{P,i} = c_i * K_{P0} \\ K_{I,i} = c_i * K_{I0} \end{cases} \tag{38}$$

where $K_{P,i}$ and $K_{I,i}$ are the PI parameters of the $i$-th WF; $c_i$ is the adaptive gain of the $i$-th WF with the value range of 0 to 1.

Since the rotor kinetic energy is supposed as the only energy source for the WF to provide FFS in this paper, the WF with more kinetic energy should undertake more frequency support tasks accordingly. Therefore, the adaptive gain is designed as

$$c_i = \frac{E_{k0,i} - E_{k,i}^{min}}{E_{k,i}^{max} - E_{k,i}^{min}} \tag{39}$$

where

$$E_{k,i} = \sum_{j=1}^{N_{WT,i}} \frac{1}{2} J_{WT,ij}\omega_{r,ij}^2 \tag{40}$$

$E_{k0,i}$ is the steady-state kinetic energy of the $i$-th WF before disturbance; $E_{k,i}^{min}$ and $E_{k,i}^{max}$ are the minimum kinetic energy and maximum kinetic energy of the $i$-th WF.

Eq. (39) shows that the more the steady-state kinetic energy of the WF, the larger the PI parameters. Correspondingly, the response power generated by the PI controller is larger, which is consistent with the design expectation. Besides, the proposed adaptive gain strategy relies only on the local information of the WF itself, requiring no extra communication among WFs.

### C. Implementation Details

#### 1) Estimation of power deficit

As shown in Fig. 6, the power deficit $P_d$ is required to be known for implementing the proposed method. Based on the measured frequency at the point of common coupling (PCC), the WF controller can estimate the magnitude of the power deficit by using the method proposed in [30]. That is

$$\hat{P}_d = -\frac{2H}{N_{WF}} * \sum_{i=1}^{N_{WF}} \frac{f_i(\Delta t + 0) - f_i(0)}{\Delta t} \tag{41}$$

where $\hat{P}_d$ denotes the estimated value of $P_d$; $f_i$ is the local frequency measured at the $i$-th WF; $\Delta t$ is the time window for calculating the RoCoF; $t = 0$ denotes the moment of the power deficit event. To balance the influence of $\Delta t$ on the RoCoF approximation accuracy and the rapidity of the start-up speed of FFS, $\Delta t$ is set to 300ms in this paper.

#### 2) Exit strategy

When completing the frequency support, all WT in the WF should recover the optimal operating point. According to the operating characteristics of the WT, MPPT control can ensure the state recovery of the WT. Therefore, a widely-used MPPT-based exit strategy is adopted [20], [21], [26]. That is, once the active power intersects the MPPT curve, the WT exits FFS and switches to MPPT control for the rotor speed recovery.

Fig. 8 shows the whole operating trajectory of the WT under the combined action of the proposed FFS and the MPPT-based exit strategy. In the FFS stage, the WT first increases the output power to mitigate the system frequency drop. Then it gradually reduces the output power to recover the KE released in the early stage. Finally, the MPPT-exit strategy is activated to ensure that the WT returns to the initial steady-state operation point.

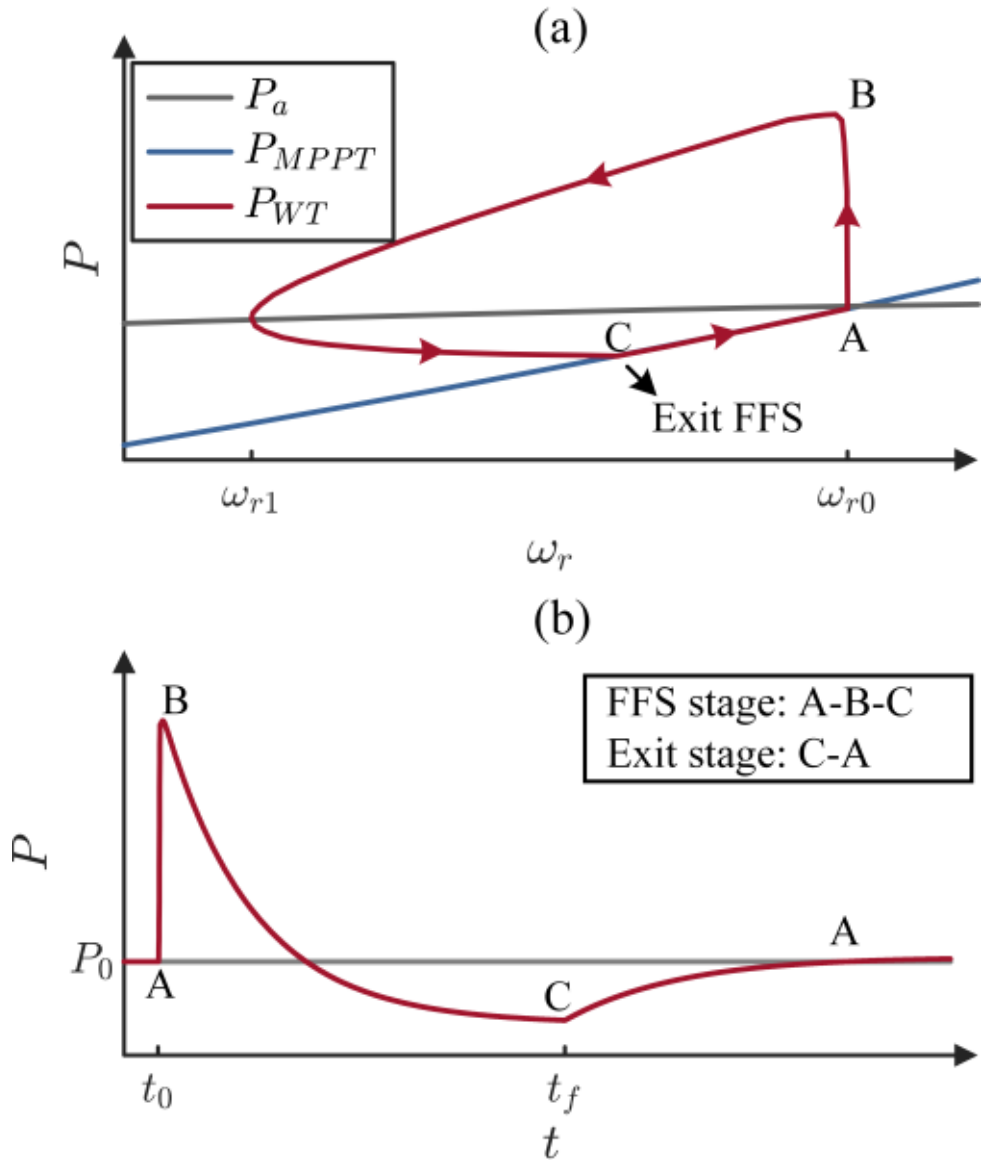

Fig.8 Operating trajectory of the WT under proposed FFS control with exit strategy. (a) Power-speed trajectory. (b) Power-time trajectory.

TABLE II
PARAMETERS OF THE WIND TURBINE

| Parameter | Value |
|---|---|
| Rated voltage (kV) | 0.69 |
| Rated capacity (MVA) | 5.556 |
| Rated active power (MW) | 5 |
| Rated wind speed (m/s) | 11.0 |
| Wind turbine diameter (m) | 112 |
| Nominal rotor speed (rpm) | 1484.153 |
| Rotor and turbine inertia (kg·m$^2$) | 1993.285 |
| Maximum rotor speed (p.u.) | 1.2 |
| Minimum rotor speed (p.u.) | 0.7 |

## V. RESULTS AND DISCUSSIONS

### A. Case Study I: A Single-WF System

A power system consisting of one SG and one WF is constructed in the DIgSILENT/PowerFactory. The total active load is 150MW, i.e. $P_L = 150\text{MW}$, with a frequency damping factor of $D_f = 1$. The rated capacity of the SG is 200MVA, with a droop factor of $R = 0.05$ and inertia time of $H = 4\text{s}$. The WF is aggregated by 20 DFIG-based WTs with parameters listed in TABLE II. The wind speed is $9\text{m/s}$. The power deficit is set to be a load surge accounting for 10% of the total active load at $t = 2\text{s}$. Based on (41), the power deficit is estimated as 14.2MW, with a relative estimation error of 5.33%. Then, the steady-state frequency excursion is calculated as $-0.169\text{Hz}$ by (4). Assuming that the maximum frequency excursion should not exceed $\pm 0.2\text{Hz}$, the value of $\alpha$ can be calculated as 1.18. According to the design principle shown in (30), PI parameters are set as $K_{P0} = 148$ and $K_{I0} = 13.5$.

#### 1) Response characteristics

The response characteristic of the proposed method is shown in Fig. 9. The results indicate that it forces the system frequency to track the optimal trajectory by regulating the output power of the WF. Once subjected to a power deficit event, the initial RoCoF of the system is estimated by the WF according to the average RoCoF in the time interval of $\Delta t$, as shown in Fig. 9(a). Correspondingly, this average-based estimation method leads to a $\Delta t$ delay in the response of FFS, as shown in Fig. 9(b). Then, the output power of the WF increases under the action of the FFS to mitigate the frequency drop. In the stage where the frequency is constant, the output power of the WF gradually decreases. Once the output power intersects with the MPPT power, the WF smoothly switches back to MPPT control, which promotes the recovery of its rotor speed and effectively avoids a secondary frequency droop. Under the MPPT control, the WF gradually returns to the optimal operating point. Accordingly, the system frequency also transits to the steady state value, thus completing the PFR.

#### 2) Adaptivity to the black-box governor model

As emphasized above, the core merit of the proposed method is its adaptivity to the black-box governor model. In this section, three types of governor are verified, including the simplified governor shown in (23) and two classical governor models, i.e. the *IEEEG1* governor and the *IEEEG3* governor, as shown in Fig. 10. The detailed parameters of the *IEEEG1* and *IEEEG3* governor are exhibited in TABLE VI in Appendix B. Although the governor models are given, they are assumed to be unknown when the proposed method is applied. The system frequency under the above three governor models is shown in Fig. 11(a). Taking the simplified governor model as an example, the adaptability of the proposed method to the governor parameter varying is verified by changing the value of $T_g$ , with results shown in Fig. 11(b). It can be seen that whether for different governor models or different governor parameters, the proposed method can regulate the output power of WFs to shape the system frequency as the optimal trajectory. Hence, the results verify its excellent adaptivity to the black-box governor.

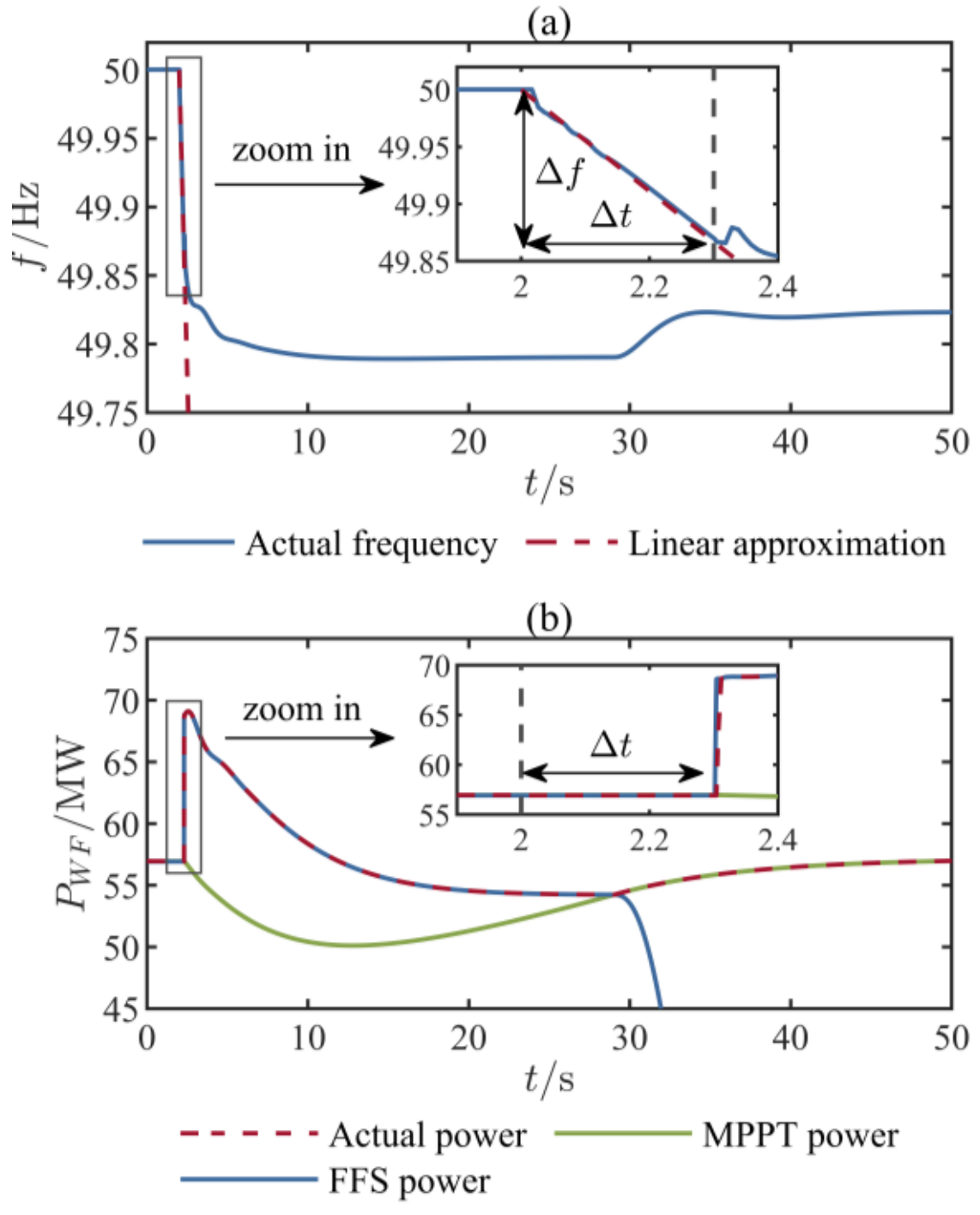

Fig. 9 Response characteristics of the proposed method.

Based on the *IEEEG1* model, the influence of the modeling error of governors on the performance of the proposed method and the model-based FFS shown in Fig. 3 is further compared, both of which aim at achieving the optimal frequency trajectory. It is assumed that there are random modeling errors of the governor and different error levels are tested. At each error level, 1,000 tests have been performed. Specifically, the minimum values of the frequency nadir are listed in TABLE III. The results indicate that the performance of the proposed method is not affected by the modeling error of governors. Admittedly, the model-based method can achieve better performance when the governor is accurately modeled. However, it is notably impacted by the modeling error level. At 10% modeling error, the worst-case frequency nadir under the model-based method is reduced to $-0.2250\text{Hz}$ , which is 12.5% lower than the expected value. Hence, compared with the model-based method, the proposed method has better robustness to the modeling error of governors.

*3) Comparison validation*

To verify the effectiveness of the proposed method, the other three FFS control strategies are performed as comparisons, namely fixed VIC in [4], adaptive VIC in [5], and SIC in [14]. Then, the performance of the above methods is tested with the frequency nadir as the evaluation index. In view of the uncertainty of actual

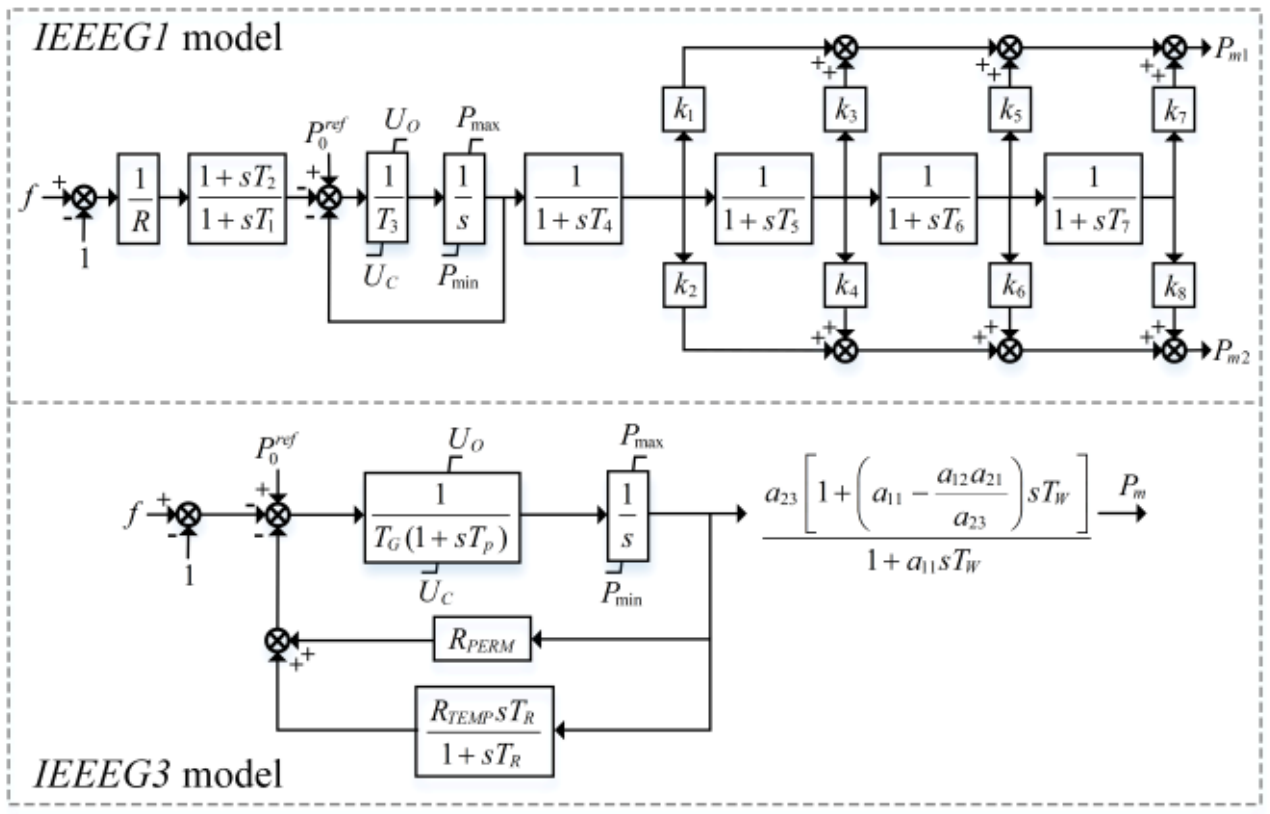

Fig. 10 *IEEEG1* and *IEEEG3* governor model.

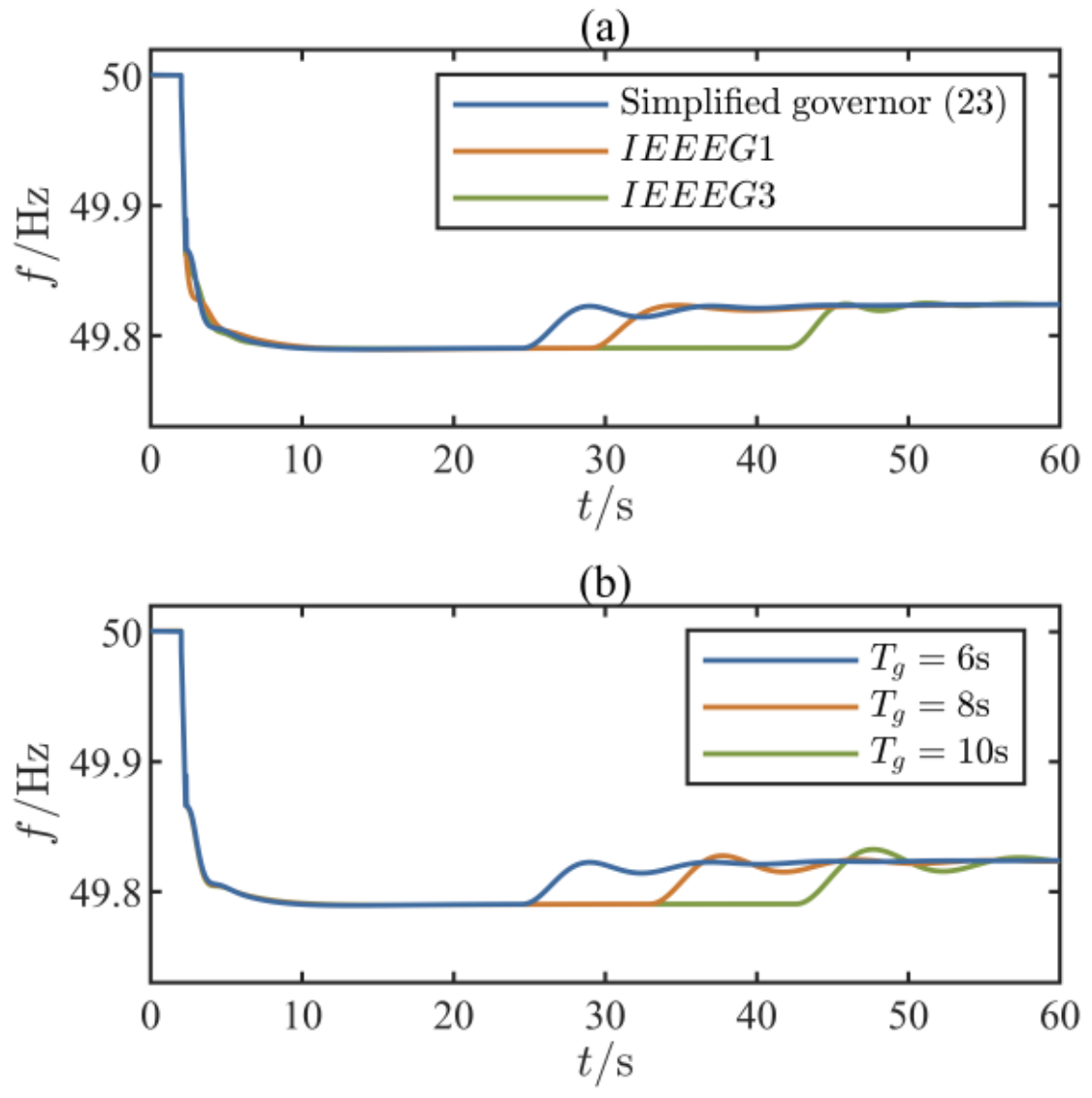

Fig. 11 Verification of model-free property of proposed control. (a) System frequency dynamics under different governor models. (b) System frequency dynamics under different values of $T_g$ .

TABLE III
COMPARISON OF FREQUENCY NADIR CONSIDERING MODELING ERRORS

| Modeling error level (%) | $\min(\Delta f_{nadir})$ /Hz | | $\frac{\min(\Delta f_{nadir}) - \Delta f_{nadir}^{ref}}{\Delta f_{nadir}^{ref}} \times \%$ | |
|---|---|---|---|---|
| | Model-based FFS | Proposed FFS | Model-based FFS | Proposed FFS |
| 2.5% | -0.2096 | -0.2108 | 4.8% | 5.4% |
| 5.0% | -0.2138 | -0.2108 | 6.9% | 5.4% |
| 7.5% | -0.2187 | -0.2108 | 9.05% | 5.4% |
| 10% | -0.2250 | -0.2108 | 12.5% | 5.4% |

Note: $\Delta f_{nadir}^{ref} = -0.2\text{Hz}$ denotes the nadir of the optimal frequency trajectory.

power deficit events, the influence of the power deficit magnitude on the frequency nadir is first tested. Generally, a 10% load surge is generally regarded as a severe power deficit event. Hence, the power deficit is set to not exceed this upper bound. The frequency nadir under a series power deficit is shown in Fig. 12(a). It can be seen that the frequency nadir under different FFS control strategies is linear with the power deficit magnitude in a certain range, and the frequency nadir of the proposed method is higher than other methods. Furthermore, the impact of system operating conditions varying on the performance of FFS control strategies is tested with the system inertia time as a representative, as shown in Fig. 12(b). The results show that the decline of system inertia will impair the performance of some FFS control strategies, such as fixed VIC and SIC. In contrast, the frequency nadir of the proposed method is almost not affected by the system inertia time, indicating that it has good robustness to the change of system conditions.

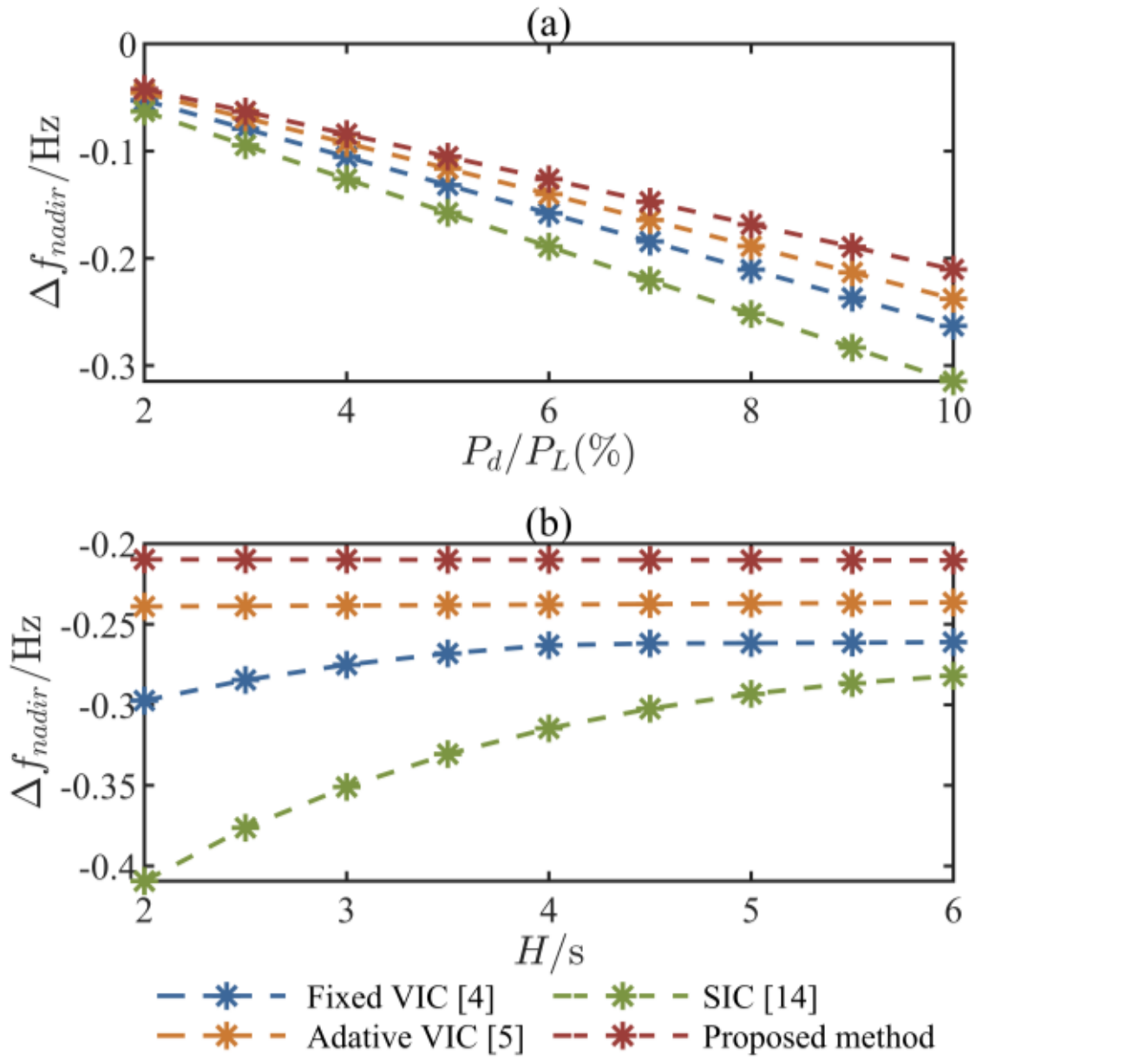

Fig. 12 Comparison of frequency nadir under different FFS control strategies. (a) Influence of power deficit magnitude. (b) Influence of system inertia time.

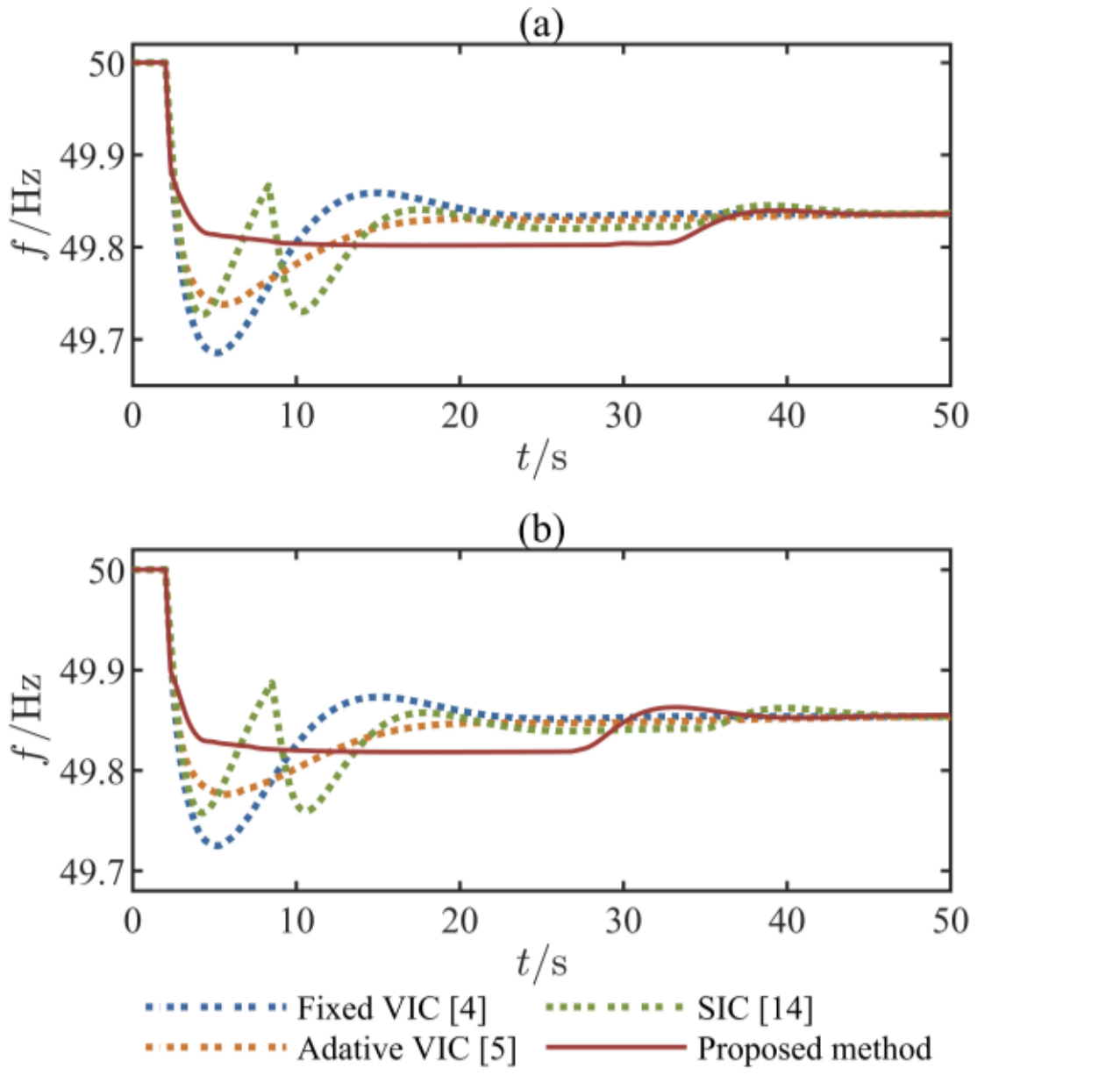

Fig. 13 System frequency trajectory under different frequency support control of WTs. (a) Load surge disturbance. (b) SG tripping disturbance.

### *B. Case Study II: A Multi-WF System*

A modified IEEE 39-bus system with 5 WFs is constructed to verify the performance of the proposed method in a multi-WF system. The structure of the IEEE 39-bus system is exhibited in [31]. Each WF is composed of 80 WTs with a rated power of 5MW, and the detailed model is the same as in *Case I*. The 5 WFs locate in Bus8, Bus14, Bus16, Bus18, Bus 22, respectively. As for the SGs, the *IEEEG1* governor is applied, with detailed parameters shown in TABLE VII in Appendix B. The initial active load is 6097.2MW with a damping factor of $D_f = 1.47$. Taking the sum of the rated capacity of all SGs as the base value, the system inertia and drooping factor are then calculated as: $H = 4.1289\text{s}$, $1/R = 17$. The potential maximum power deficit is assumed as 500MW, accounting for about 8.2% of the total active load. Then, the steady-state frequency excursion is calculated as $-0.163\text{Hz}$ under this disturbance. To ensure the frequency nadir be within $\pm 0.2\text{Hz}$, the value of $\alpha$ is calculated as 1.226. Based on (30), the PI parameters are set as $K_{P0} = 120$ and $K_{I0} = 11.5$.

*1) Different types of power deficit event*

Given the various types of power deficit events in the large-scale power system, the performance of the proposed method under the following two typical power deficit events is tested.

**Load surge**: Load16 increases 500MW at $t = 2\text{s}$.

**Generator tripping**: SG G5 trips at $t = 2\text{s}$.

The system frequency trajectories under the above two power deficit events are shown in Fig. 13, and the maximum frequency excursions are listed in TABLE IV. The results indicate that the proposed method effectively improves the frequency nadir in different types of power deficits. Compared with fixed VIC and adaptive VIC, the maximum frequency excursion is improved by 37.01% and 24.39% in the event of load surge, and 33.93% and 18.92% in the event of generator tripping. The performance improvement is stimulated by breaking the confine of imitating the frequency response characteristics of SG but collaborating with SGs. Compared with VIC, the proposed method not only improves the frequency nadir but also avoids the problem of secondary frequency drop.

TABLE IV
COMPARISON OF MAXIMUM FREQUENCY EXCURSION

| | Load surge | Generator tripping |
|---|---|---|
| Fixed VIC [4] | -0.3145Hz | -0.2750Hz |
| Adaptive VIC [5] | -0.2620Hz | -0.2241Hz |
| SIC [14] | -0.2732Hz | -0.2422Hz |
| Proposed control | -0.1981Hz | -0.1817Hz |

*2) Different wind speed distribution*

Given the influence factors such as terrain and atmosphere, the wind speed of WFs may be quite different. In this section, the performance of the proposed method is verified when there is a wind speed difference in the multi-WF system. Firstly, maintaining the wind speed of WF2-5 at 9m/s, the frequency support dynamics of WF1 under a series of wind speeds are tested. As shown in Fig. 14, with the decrease in the wind speed, the adaptive gain coefficient $c$ becomes smaller based on (39). Then, WF1 does generate less support power, and the rotor speed drop is significantly alleviated. That is, WF1 takes less responsibility for system frequency support as the wind speed decreases, which proves the feasibility of adjusting the WF's frequency response by configuring PI gains.

Based on the above results, the performance of the proposed adaptive gain strategy of the PI parameters for the multi-WF coordination is further verified. Specifically, the wind speed of the five WFs is set as 6.5m/s, 7.5m/s, 8.5m/s, 9.5m/s, and

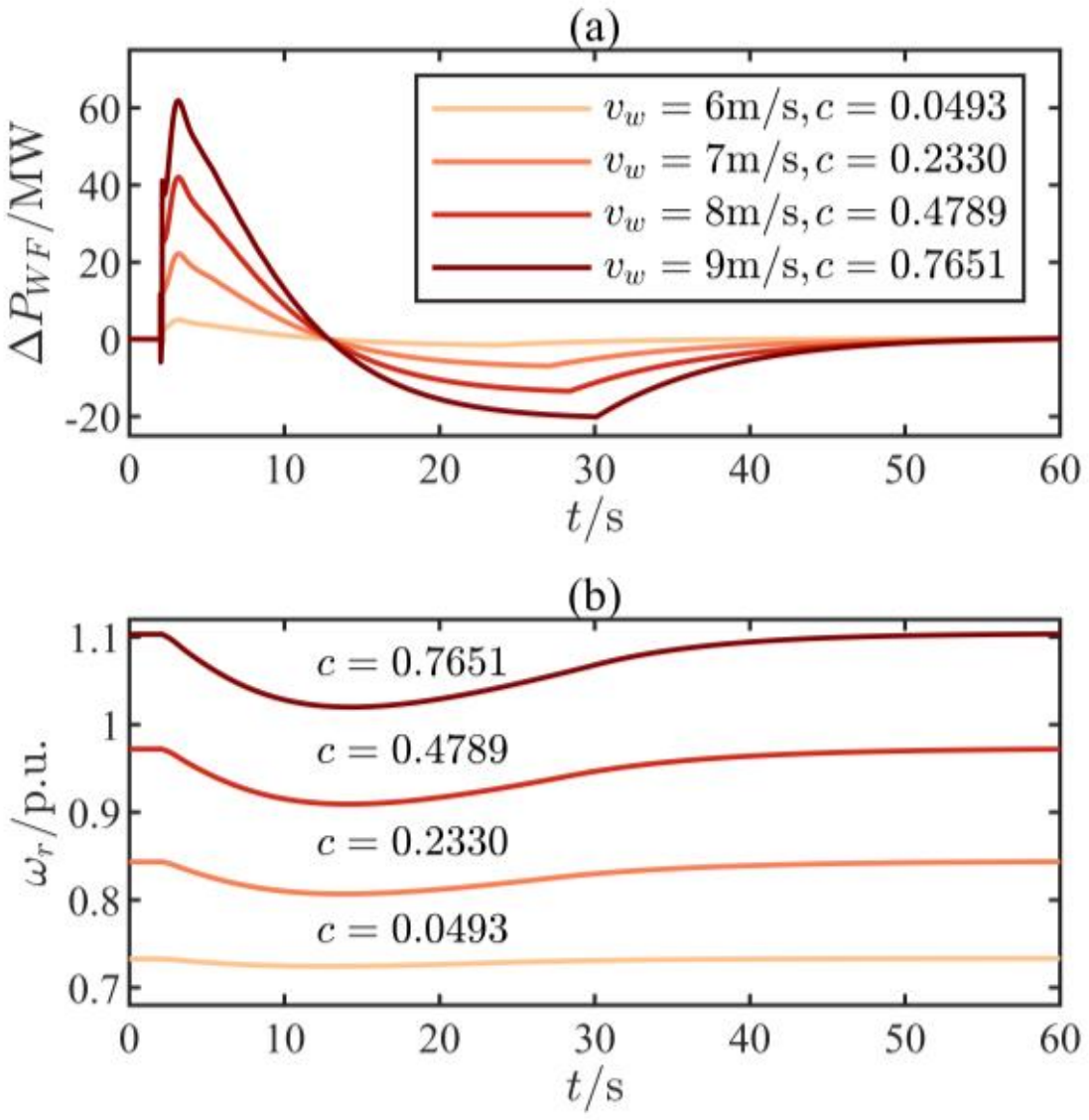

Fig. 14 Frequency support dynamics of WF1 under different wind speeds.

TABLE V
STEADY-STATE OPERATING PARAMETERS OF WFS

| | WF1 | WF2 | WF3 | WF4 | WF5 |
|---|---|---|---|---|---|
| $v_w$(m/s) | 6.5 | 7.5 | 8.5 | 9.5 | 10.5 |
| $\omega_{r0}$(p.u.) | 0.7852 | 0.9063 | 1.0387 | 1.1619 | 1.2 |
| $c$ | 0.1332 | 0.3488 | 0.6199 | 0.9053 | 1.0 |

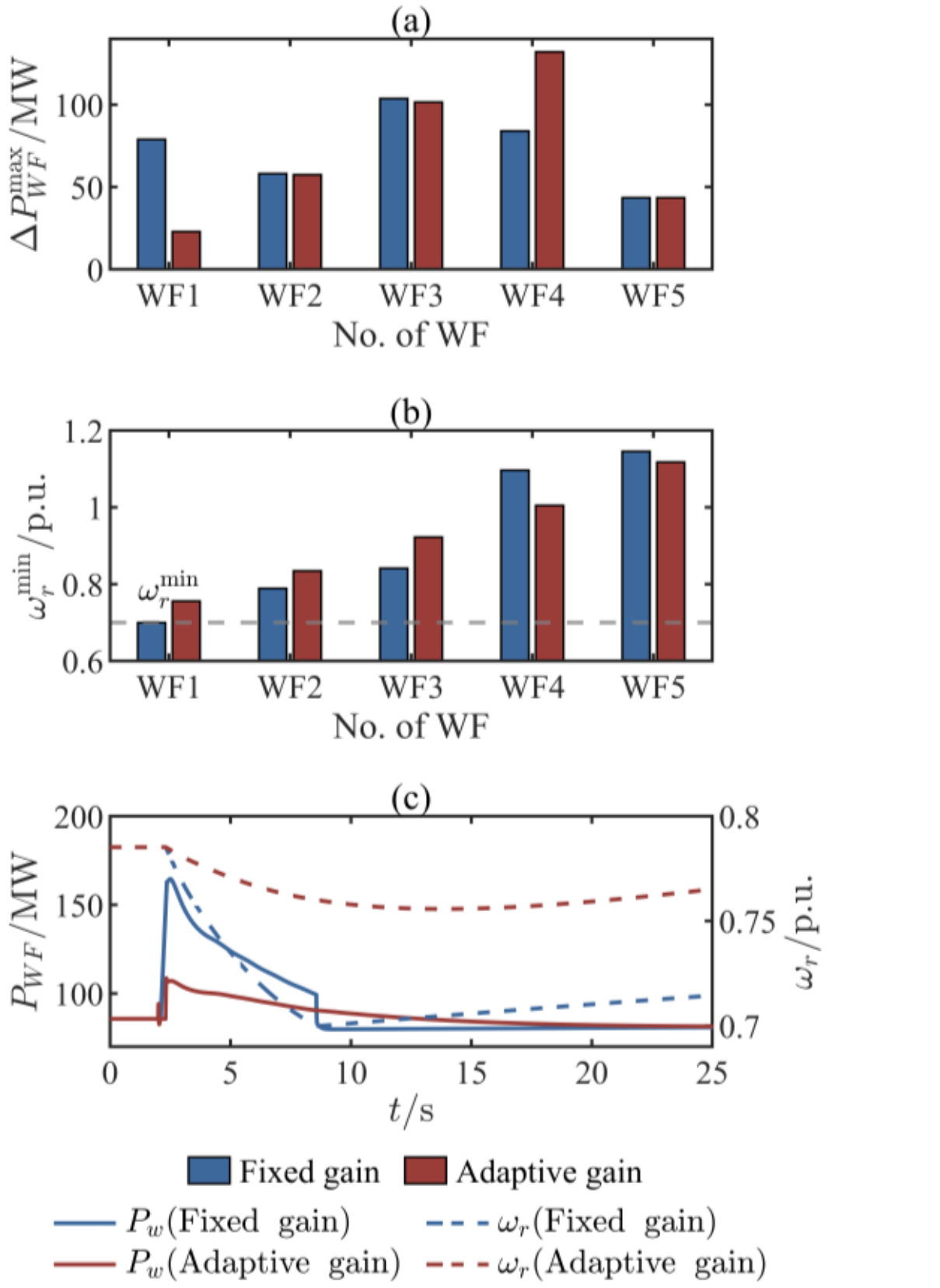

Fig. 15 Comparison of fixed gain and proposed adaptive gain strategies. (a) Maximum support power of WFs. (b) Minimum rotor speed of WFs. (c) Output power and rotor speed dynamics of WF1.

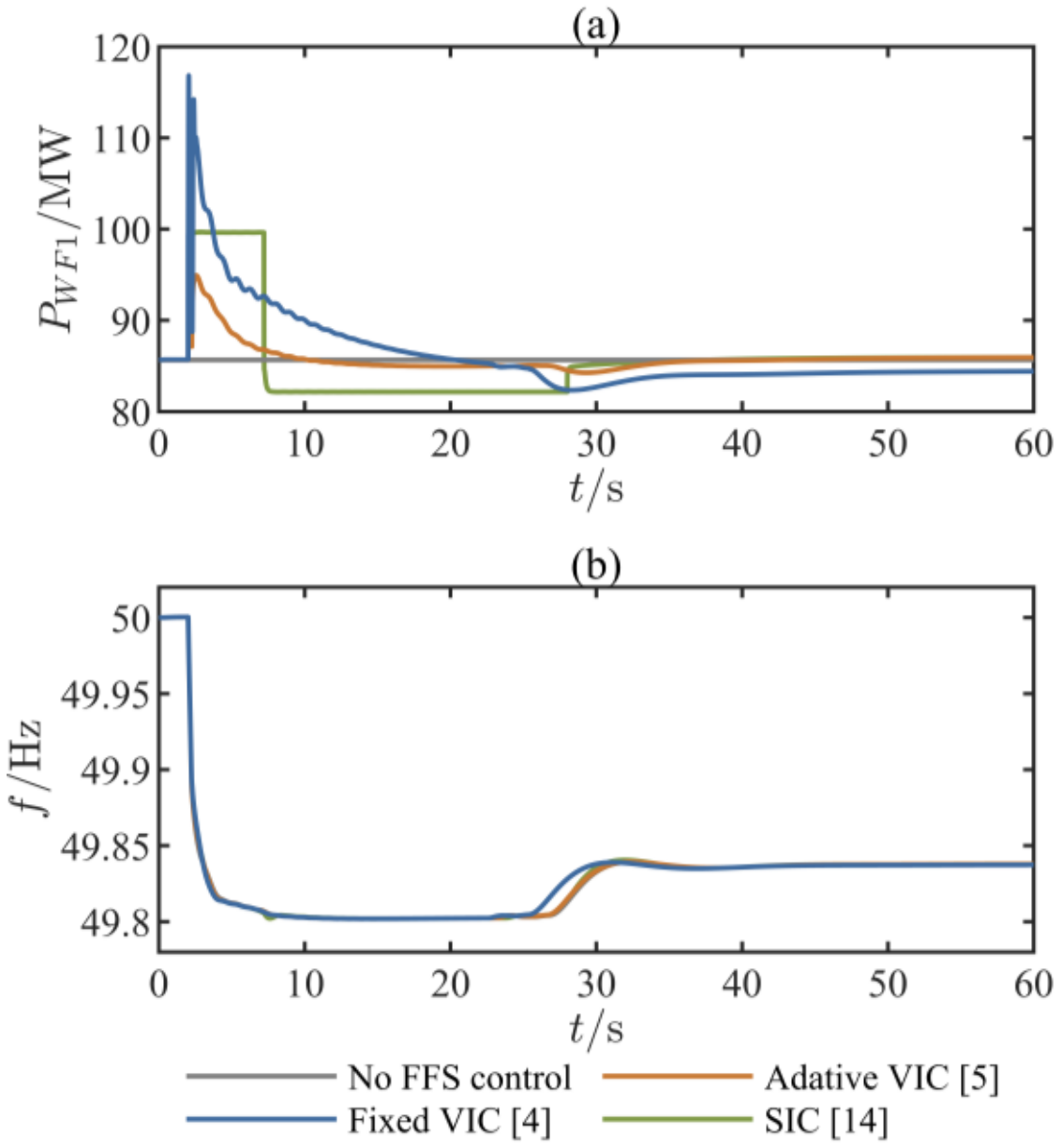

Fig. 16 Compatibility of the proposed FFS to other control strategies. (a) Output power of WF1 under different FFS strategies. (b) System frequency trajectory.

10.5m/s, respectively. Then, the adaptive gain $c$ of each WF can be calculated, listed in TABLE V. Meanwhile, the fixed gain ($c = 1$ for all WFs) is performed as a comparison. As shown in Fig. 15, all WFs are equally responsible for the FFS task under the fixed gain strategy. For the WF operating at a low wind speed, such as WF1, it faces the risk of over-deceleration to touch the lower bound. The support power and rotor speed dynamics of WF1 are detailedly shown in Fig. 15(c). The results show that the rotor speed of WF1 with fixed gain drops rapidly to the lower bound due to excessive frequency support, which increases the risk of stall. In contrast, the proposed method limits the support power of low-wind speed WF by reducing the PI parameters, thereby significantly reducing its over-limit risk in the rotor speed. Since the total FFS task is determined, more FFS task is undertaken by the high-wind speed WF. Hence, the proposed method makes the FFS task reasonably shared among WFs.

*3) Compatibility with other FFS control strategies*

Considering a more general situation, the proposed PI-based FFS may not be adopted by all WFs. Specifically, the WF1 is supposed to adopt different FFS strategies and other WFs remain the proposed FFS strategy. Fig.16 shows that when WF1 presents various output characteristics, the proposed FFS can still force the remaining WFs to track the optimal frequency trajectory. This means the proposed FFS has good compatibility with other FFS strategies. That is, the desired control objectives can still be achieved without forcing all WFs to execute the proposed PI-based FFS.

## *C. Discussion*

The above results verify the comprehensive performance of the proposed method. However, it must be admitted that the frequency trajectory optimization-oriented FFS of WFs is in its infancy, and its feasibility in the actual power system requires further research. Compared with the most popular VIC scheme, the proposed method faces the following challenges.

1) **Dependence on system parameters**. Despite getting rid of modeling the governor dynamics, it still relies on some system parameters such as system inertia and power deficit, et al. Once these parameters cannot be obtained, it will inevitably impede the application of the proposed method. Besides, the system parameter dependency will increase the implementation cost as additional measurement and communication are needed.

2) **Requirement on wind power scale**. To shape the system frequency into the desired optimal trajectory, the wind power scale should be large to provide sufficient KE for frequency support. So, the proposed method cannot achieve the desired control effect in the scenario with a small scale of wind power.

Fortunately, parameter estimation is always a vital research topic. A lot of effective solutions have been proposed in the existing research, such as power deficit estimation in [32] and inertia estimation in [33], which provide potential solutions for the proposed method. Meanwhile, as the scale of converter-type power sources increases, the high controllability and enormous frequency support potential of the source side enhance the plasticity of the system frequency trajectory. Given the pros and cons of the proposed method, further research can be carried out including but not limited to the following three aspects.

1) **Reducing dependence on system parameters.** For large-scale power systems, it is not always easy to accurately obtain system parameters. Hence, reducing the dependence on system parameters is of great significance for the application of the method in the actual power system.

2) **Coordinating with other FFS controls.** It is attractive to develop an event-driven coordination strategy so that the proposed method can work in severe scenarios while maintaining the easy-to-implement FFS control in non-severe scenarios. In this way, both the performance and the cost of FFS of WFs can be considered to optimize the frequency dynamics.

3) **Developing other flexible frequency regulation sources**. In addition to WTs, other flexible frequency regulation sources such as photovoltaic and energy storage can be considered in FFS. Studying the implementation of the optimal frequency trajectory-oriented FFS schemes in these flexible frequency regulation sources will facilitate expanding the application scope of the method.

## VI. Conclusion

In this paper, a PI-based FFS of WFs has been proposed to improve the transient frequency safety of the power system. Benefiting from the model-free property of the PI controller, the proposed method can achieve the optimal frequency trajectory without relying on the governor model. The adaptivity to the black-box governor model greatly improves the application potential in the actual power system as the arduous work of governor modeling is completely avoided. Besides, the adaptive gain strategy achieves multi-WF coordination without extra communication demands. Notably, the proposed FFS has good compatibility with other FFS strategies and can achieve the optimal frequency trajectory without being forced to install on all WFs. Finally, simulation results verify that the proposed method is simple to implement but can effectively improve the transient frequency safety of power systems.

## Appendix A

## Derivation of Value Range of PI Parameters

According to (25), it is easy to know that $b_{num}$ and $a_{den}$ are larger than 0; $b_{den}$ is generally less than 0 as $\omega \le \omega_{up}^{\max}$ is very small; and the sign of $a_{num}$ is uncertain.

Then, according to the sign of $b_{num}$ and $b_{den}$ and the quantitative relation shown in (28), one obtains

$$\left|b_{den}\right| = -b_{den} \ge 10 b_{num} \Rightarrow K_I \ge 9h(\omega,T_g) - 2H\omega^2 \tag{42}$$

where $h(\omega,T_g) = \dfrac{\omega^2}{R\left[1/T_g + \omega^2 T_g\right]}$.

As $\omega \ge 0$, the above inequality can be further scaled

$$K_I \ge 9h(\omega,T_g) \ge 9h(\omega,T_g) - 2H\omega^2,\ \omega \ge 0 \tag{43}$$

The denominator of $h(\omega,T_g)$ reaches its minimum value at $T_g = 1/\omega$, which makes $h(\omega,T_g)$ reach its maximum value, i.e.

$$h(\omega,T_g) \le h(\omega,T_g = \frac{1}{\omega}) = h_1(\omega) = \frac{\omega}{2R} \tag{44}$$

Obvious, $h_1(\omega)$ is a monotonically increasing function of $\omega$, which takes its maximum value when $\omega$ takes its maximum value $\omega_{up}^{\max}$. For the maximum value of $h_1(\omega)$, (43) should still hold. Then, the value range of $K_I$ can be derived as

$$K_I \ge 9h_1(\omega_{up}^{\max}) \tag{45}$$

As for the value range of $K_P$, it should be discussed in two scenarios because the sign of $a_{num}$ is uncertain. Before that, the scale is first performed on $a_{den}$, as follows.

$$\left|a_{den}\right| = a_{den} = K_P + D_f + g(\omega,T_g) \ge K_P \tag{46}$$

where $g(\omega,T_g) = \dfrac{1}{R\left[1+\left(\omega T_g\right)^2\right]}$, which will monotonically decrease with respect to $\omega$ and $T_g$ when $\omega \ge 0$ and $T_g \ge 0$.

In the scenario of $a_{num} \ge 0$, one obtains

$$\left|a_{den}\right| \ge K_P \ge 10\left|a_{num}\right| = 10\left(K_g^* - D_f - g(\omega,T_g)\right) \tag{47}$$

Based on the monotonicity of $g(\omega,T_g)$, Eq. (47) can be further expressed as

$$K_P \ge 10\left(K_g^* - D_f - g(\omega_{up}^{\max}, T_g^{\max})\right) \tag{48}$$

Similarly, the value range of $K_P$ can be derived as (49) in the scenario of $a_{num} < 0$.

$$K_P \ge 10\left(D_f + g(0,0) - K_g^*\right) = 10\left(K_g - K_g^*\right) \tag{49}$$

Then, the value range of $K_P$ should be the intersection of (48) and (49).

## Appendix B

## Parameters of Governors

TABLE VI
Governor Parameters Setting in the Single-WF Case

| IEEEG1 | $S$ | $H$ | $1/R$ | $K_1$ | $K_3$ | $K_5$ | $K_7$ |
|---|---|---|---|---|---|---|---|
| | $T_1$ | $T_3$ | $T_4$ | $T_5$ | $T_6$ | $T_7$ | |
| SG | 200 | 4 | 20 | 0.3 | 0.15 | 0.3 | 0.25 |
| | 0.2 | 0.1 | 0.25 | 3 | 3.5 | 0.25 | |
| IEEEG3 | $S$ | $H$ | $1/R_P$ | $R_T$ | $T_G$ | $T_P$ | $T_R$ |
| | $T_W$ | $\alpha_{11}$ | $\alpha_{13}$ | $\alpha_{21}$ | $\alpha_{23}$ | | |
| SG | 200 | 4 | 20 | 0.2 | 0.05 | 0.04 | 3 |

| | | | | | |
|---|---|---|---|---|---|
| | 0.75 | 0.5 | 1 | 1.5 | 1 |

TABLE VII
GOVERNOR PARAMETERS SETTING IN THE MULTI-WF CASE

| *IEEEG1* | $S$ | $H$ | $1/R$ | $K_1$ | $K_3$ | $K_5$ | $K_7$ |
|---|---|---|---|---|---|---|---|
| | $T_1$ | $T_3$ | $T_4$ | $T_5$ | $T_6$ | $T_7$ | |
| G1 | 1200 | 5 | 17 | 0.2 | 0.2 | 0.35 | 0.25 |
| | 0.2 | 0.2 | 0.4 | 5 | 6 | 0.4 | |
| G2 | 700 | 4.329 | 17 | 0.3 | 0.32 | 0.18 | 0.2 |
| | 0.1 | 0.2 | 0.3 | 4 | 4.5 | 0.4 | |
| G3 | 800 | 4.475 | 17 | 0.22 | 0.22 | 0.3 | 0.26 |
| | 0.15 | 0.1 | 0.4 | 5.5 | 5 | 0.4 | |
| G4 | 800 | 3.575 | 17 | 0.26 | 0.28 | 0.3 | 0.16 |
| | 0.1 | 0.1 | 0.2 | 4 | 4.5 | 0.4 | |
| G5 | 600 | 4.333 | 17 | 0.25 | 0.3 | 0.3 | 0.15 |
| | 0.15 | 0.15 | 0.4 | 5 | 4 | 0.5 | |
| G6 | 800 | 4.35 | 17 | 0.2 | 0.3 | 0.3 | 0.2 |
| | 0.3 | 0.2 | 0.3 | 4.5 | 4.5 | 0.4 | |
| G7 | 700 | 3.771 | 17 | 0.25 | 0.25 | 0.3 | 0.2 |
| | 0.25 | 0.1 | 0.1 | 4.5 | 4 | 0.4 | |
| G8 | 700 | 3.471 | 17 | 0.2 | 0.25 | 0.35 | 0.2 |
| | 0.1 | 0.15 | 0.3 | 5.5 | 5 | 0.5 | |
| G9 | 1000 | 3.45 | 17 | 0.2 | 0.25 | 0.35 | 0.2 |
| | 0.3 | 0.25 | 0.4 | 5 | 4 | 0.5 | |
| G10 | 1000 | 4.2 | 17 | 0.25 | 0.3 | 0.25 | 0.2 |
| | 0.2 | 0.15 | 0.3 | 4 | 4 | 0.4 | |

Note: Other parameters of the *IEEEG1* model are given as follows, $U_O = 0.3$, $U_C = -0.3$, $P_{\min} = 0$, $P_{\max} = 1$, $T_2 = 0$, $K_2 = K_4 = K_6 = K_8 = 0$.

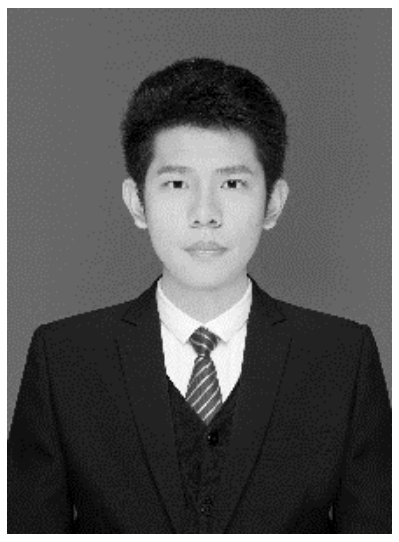

**Yubo Zhang.** (S'21) received the B.S. degree in electrical engineering from Xi'an Jiaotong University, China, in 2019. He is currently pursuing the Ph.D. degree in the school of Electrical Engineering in Xi'an Jiaotong University. His main field of interest includes the control for renewable energy, and power system frequency stability.

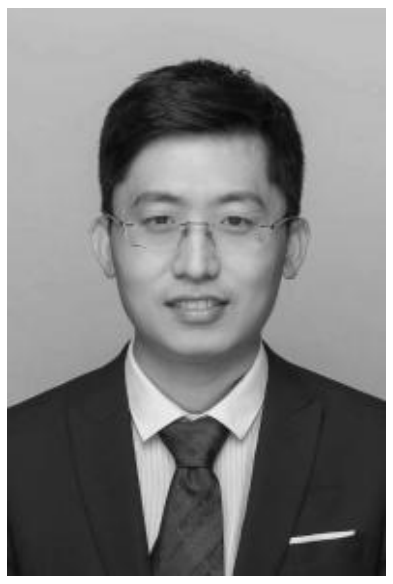

**Songhao Yang.** (S'18-M'19) was born in Shandong, China, in 1989. He received the B.S. and Ph.D. degrees in electrical engineering from the Xi'an Jiaotong University, Xi'an, China, in 2012 and 2019, respectively. Besides, he received the Ph.D. degree in electrical and electronic engineering from Tokushima University, Japan, in 2019.

Currently, he is an Associate Professor at Xi'an Jiaotong University. His research interest includes power system stability analysis and control.

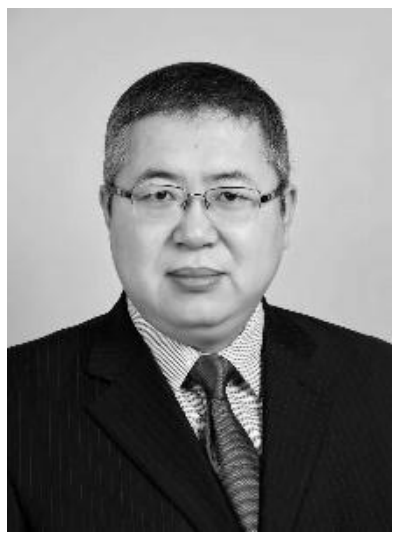

**Zhiguo Hao.** (M'10-SM'23) was born in Ordos, China, in 1976. He received his B.Sc. and Ph.D. degrees in electrical engineering from Xi'an Jiaotong University, Xi'an, China, in 1998 and 2007, respectively. Currently, he is a Professor with the Electrical Engineering Department, Xi'an Jiaotong University. His research interest includes power system protection and control.

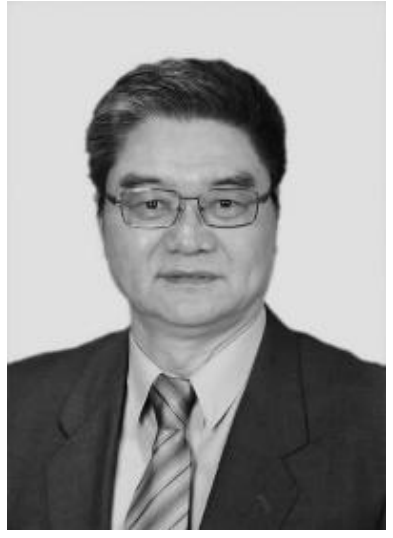

**Baohui Zhang.** (SM'99-'F'19) was born in Hebei Province, China, in 1953. He received the M.Eng. and Ph.D. degrees in electrical engineering from Xi'an Jiaotong University, Xi'an, China, in 1982 and 1988, respectively. He has been a Professor in the Electrical Engineering Department at Xi'an Jiaotong University since 1992. His research interests are system analysis, control, communication, and protection.